\documentclass[aps,preprint]{revtex4}
\usepackage{amsfonts}
\usepackage{amsmath}
\usepackage{amssymb}
\usepackage{graphicx}
\usepackage{subfigure}
\usepackage{mathrsfs, mathtools}
\usepackage{pdfpages}
\usepackage{float}
\usepackage{color}
\usepackage{xcolor}
\usepackage{caption}
\usepackage{lineno,hyperref}
\hypersetup{colorlinks =true, allcolors = blue}
\providecommand{\U}[1]{\protect\rule{.1in}{.1in}}

\begin{document}
\preprint{HEP/123-qed}
\title{ The Fermionic greybody factor and quasinormal modes of hairy black holes, as well as Hawking radiation's power spectrum and sparsity}
\author{Ahmad Al-Badawi}
\email{ahmadbadawi@ahu.edu.jo}
\affiliation{Department of Physics, Al-Hussein Bin Talal University, P. O. Box: 20, 71111,
Ma'an, Jordan.}
\author{Sohan Kumar Jha}
\email{sohan00slg@gmail.com}
\affiliation{Department of Physics, Chandernagore College, Chandernagore, Hooghly, West Bengal, India}
\author{Anisur Rahaman}
\email{manisurn@gmail.com} \affiliation{Department of Physics,
Durgapur Government College, Durgapur, Burdwan 713214, West
Bengal, India.}
\author{}
\affiliation{}


\begin{abstract}
A hairy black hole (HBH) emerges due to matter surrounding the Schwarzschild metric
when using the Extended Gravitational Decoupling (GD) approach. The fermionic greybody
factors (GFs) and quasinormal modes
(QNMs) as well as Hawking spectra and sparsity of HBH solutions are investigated. We consider massive and massless spin- 1/2 fermions, along with massless spin- 3/2 fermions. The equations of the effective
potential for fermions with different spins are derived in HBH spacetime. Then,
the rigorous bound method is used to calculate the fermionic spin- 1/2 and spin- 3/2 GFs. With the time domain integration method at our disposal, we illustrate the impact of additional parameters on the ringdown waveform of the massless fermionic spin -1/2 and spin -3/2 fields and, in turn, on their quasinormal modes. We then delve into investigating the Hawking spectra and sparsity of the radiation emitted by an HBH. Hairy parameters significantly affect the sparsity of Hawking radiation as well. We observe that the total power emitted by the BH increases both with $\alpha$ and $Q$ but decreases with $l_{0}$.   Our study conclusively shows the significant impact of the additional parameters on important astrophysical phenomena such as quasinormal modes, Hawking spectra, and sparsity. 

\end{abstract}
\volumeyear{ }
\eid{ }
\date{\today}
\received{}

\maketitle
\tableofcontents
\section{Introduction}
A fascinating prediction of Einstein's theory of general
relativity (GR) is the black hole (BH) that is entirely characterized by three
externally observed classical parameters like mass $M$, charge $Q$
and angular momentum $L$ \cite{HAW1}. Other attributes of the
BHs are exclusively determined only by these three
parameters. Information regarding matter going into or creating
the BH disappears behind its event horizon once it has
settled, making it invisible forever to outside observers.
According to the no-hair theorem, a BH
should not carry charges other than $M$, $J$, and $Q$. In
\cite{HAW2}, however, it was conjectured that additional charges
associated with inner gauge symmetries might exist in the BHs, and it is now known that BHs could have soft quantum
hair. Long-term research has examined a variety of scenarios and
potential conditions for getting around the no-go theorem
\cite{PTS, BABI, PTS1, RADU, SALGA, RBENK, ANTO, ANAB, ACIST1,
ACISR2, SVOLKOV, KAN1, KAN2, VOLKOV, CMART, KGZ}. Ref.
\cite{PTS3} has drawn much attention to studying a
fundamental scalar field. The existence of new fundamental fields
that influence the structure of the BH could lead to hairy
BH solutions. Instead of examining particular fundamental
fields to produce hair in BH solution. Instead of
considering specific fundamental fields to generate hair in BHs, in the paper \cite{olliv}, the authors assumed the presence
of a generic source in addition to the one generating the vacuum
Schwarzschild geometry. This generic source is described by a
conserved energy-momentum tensor and adopted the so-called
\cite{olliv, OVALLE0, OVALLE00, OVALLE1, OVALLE2}. Einstein's
theory of GR is a well-known and reliable
theory; nevertheless, there lies a great deal of uncertainty in
estimating the precise measurement of mass and angular momentum of
the resulting BHs. So, alternative theories of gravity
have received considerable scope to materialize \cite{ALTG}. In
this respect, studying different aspects of hairy BHs has
attracted significant attention.

The existence of gravitational waves originating from the merger of
two BHs \cite{ALTG, MERGE1, MERGE2, MERGE4, MERGE5,
MERGE6} has been confirmed from recent observation
\cite{LIGO1, LIGO2}, confirming the prediction made by the theory of GR. From an
observational perspective, the most significant phase of
gravitational wave emission can be explained in terms of proper
oscillation frequencies of the BHs, which are known as
quasinormal modes (QNMs) \cite{QNM1, QNM2, QNM3}. Press
\cite{PRESS} first used the term QNMs, but Vishveshwara
\cite{VISH} initially identified them in the simulations of
gravitational wave scattering off a Schwarzschild BH.

Perturbation of spacetime using a field is a useful tool to accumulate
precise information about the interior of the BHs. The study of
spacetime perturbation associated with various BHs using
different probes (field) has become an intriguing area of
investigation after the confirmation of gravitational wave
detection from the observation \cite{LIGO1, LIGO2}. There are
various ways to introduce perturbation, including
scalar, electromagnetic, and fermion fields. Many studies have
been conducted to identify QNMs
using perturbation through different fields against diverse geometrical backgrounds.
Scalar and electromagnetic perturbation are often used in research
to find QNMs in different geometrical environments. Despite the
various scientific efforts, there needs to be more
substantial literature on QNMs studied through perturbation using
a fermionic field as a probe. Perturbation involving the fermion field
entails an additional level of complexity due to the presence of
positive and negative energy solutions offered by the Dirac equation,
but interestingly, research has shown that at least when it comes
to Schwarzschild background, both positive and negative energy
solutions produce identical QNMs. There are numerous applications
of the quasinormal modes in GR. The study of QNMs is
crucial to analyzing the classical stability of BHs
against matter fields that are used as perturbation probes. QNMs
are essential for the AdS/CFT correspondence because they define
the relaxation durations of dual-field theories \cite{AHAR, HORO}.
For quark-gluon plasmas \cite{QGP1, QGP2}, the QNMs of
asymptotically AdS BHs play a crucial role in the
holographic description. Another remarkable possibility is raised
through the fascinating conjecture made by Hod concerning quantizing the area of BHs \cite{SHOD, OD, MAGGI}. It
is also considered how QNMs and Hawking radiation
are related \cite{KONO, KIEF}. The advancements in experimental
astrophysics and the detection of gravitational waves present a
fresh opportunity to apply the QNM approach for verifying certain
general relativity conjectures or estimating various properties of
compact sources of gravitational field \cite{BERTI}.

The perturbation of the hairy BH background is interesting in its
own right. Here, we find an extra hair (charge). This extra charge
generated here using the ingenious technique of decoupling of
energy-momentum tensor evades no-go theorem \cite{OVALLE0,
OVALLE00}. On the other hand, the perturbation of spacetime background
using a fermion field \cite{tt1,ahmad1,ahmad2} is scanty compared to the perturbation with
scalar and electromagnetic fields. Therefore, studying fermionic
greybody factors (GFs) and QNM for a hairy BH using
fermion field perturbation is instructive and of interest. This paper focuses on one recent solution reported in
\cite{olliv} among all the possible hairy BHs in literature. The
authors of Ref. \cite{hb10} study the linear stability of a BH
with scalar hair under axial gravitational perturbations and find
that the BH is linearly stable under axial perturbations. The QNM
of Hairy BH (HBH) caused by gravitational decoupling has been
studied recently in \cite{hb11}. They conclude that for HBH, the
effects of these hairy parameters on time-domain profiles and QNM
frequencies under perturbations show similar behavior. The main objectives were to analyze a
scalar perturbation in the HBH background solution and compare
it to the ordinary Schwarzschild background solution were the main
objectives \cite{hb12}. Furthermore, the HBH solution has stimulated
further research in its generalization to hairy Kerr
\cite{hb12b}.

The organization of the remaining part of the paper is as follows.
Sect. II briefly reviewed the spacetime of HBH caused by GD. In
Sect. III, we also reviewed the equations of motion related to the
spinorial wave equations, namely the Dirac and Rarita-Schwinger,
around the HBH spacetime. Further, we obtain the corresponding
effective potentials for each field, respectively. Sect. IV is
devoted to calculating the bounds of the GFs of BH and analyzing
their graphical behavior. We obtain the GFs for massive spin- 1/2
fermions, massless spin- 1/2 fermions, and massless spin- 3/2
fermions, respectively. In Sect. V, the time-domain profiles of massless fermionic spin- 1/2 and spin- 3/2 fields perturbations in HBH spacetime are given. In Sect. VI, we examine the Hawking spectra and sparsity of the radiation emitted by HBHs. The
conclusion is given in Sect. VII.

\section{Brief review of HBH}
In this section, we briefly describe the HBHs obtained by the GD
in Ref. \cite{olliv} through the Minimal Geometric Deformation
(MGD) extended (for details about GD and MGD, see
\cite{hb1,hb2,hb3,hb4,hb5,hb6,hb7,hb8}). The HBH spacetime is
given by
\begin{equation}
ds^{2}=g(r)dt^{2}-\frac{1}{g(r)}dr^{2}-r^{2}\left( d\theta ^{2}+\sin
^{2}\theta d\phi ^{2}\right) \label{M1}
\end{equation}%
where
\begin{equation}
g(r) =1-\frac{2M}{r}+\frac{Q^{2}}{r^{2}}-\alpha \left( \frac{
l_{0}+Me^{-r/M}}{r}\right) \label{mf1},
\end{equation}
in which, $M$ is the mass and $\left( \alpha ,Q,l_{0}\right) $ are the GD
HBH parameters. We note that $Q$ is not necessarily the electric
charge, but could be a tidal charge of extra-dimensional origin or any other
charge for the Maxwell tensor. This solution is presented by using GD and
dominant energy condition (DEC). It is called the \textquotedblleft
charged\textquotedblright\ HBH which extends a Reissner-Nordstr\"{o}m-like
metric. It is readily seen that for $\alpha =0$ and $Q=0$, the metric (\ref%
{M1}) reduces to the Schwarzschild BH. The event horizon is given by the
solution of
\begin{equation}
\alpha l_{0}=r_{h}-2M+\frac{Q^{2}}{r_{h}}-\alpha Me^{-r_{h}/M}. \label{hor1}
\end{equation}%
For DEC to be fulfilled, $r_{h}\geq 2M$, which results in additional
restrictions to $Q$ and $l_{0}$, namely%
\begin{equation}
Q^{2}\geq 4\alpha \left( M/e\right) ^{2}\text{ \ and \ }l_{0}\geq M/e^{2}.
\label{con1}
\end{equation}%
We see that from Eq. (\ref{con1}) the case of $Q=0$ and $\alpha $ non-zero
is not allowed. Consequently, this case is not discussed in this paper.

\begin{figure}[H]
\centering
\subfigure[]{
\label{pls}
\includegraphics[width=0.6\columnwidth]{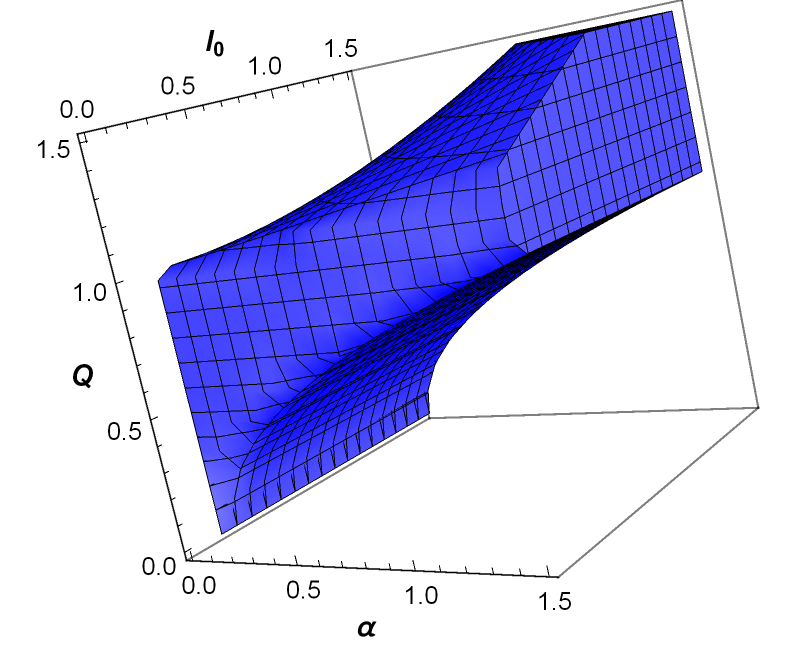}
}
\caption{Parameter space for the existence of an HBH.}
\label{ps}
\end{figure}

Throughout this work, we will focus on massive BH $(e^{-r/M}<<1)$. As there is no analytical
solution to Eq. (\ref{hor1}), we can expand the metric function Eq. (\ref
{mf1}) as
\begin{equation}
g(r)\simeq 1-\frac{2M}{r}+\frac{Q^{2}}{r^{2}}-\frac{\alpha \left(
l_{0}+M-r\right) }{r}. \label{mf2}
\end{equation}%
A calculation of the horizon radius can be done as follows:
\begin{equation}
r_{\pm }=\frac{2M+\alpha l_{0}+\alpha M\pm \sqrt{\left( 2M+\alpha M+\alpha
l_{0}\right) ^{2}-4\left( 1+\alpha \right) Q^{2}}}{2(1+\alpha )}.
\label{hor2}
\end{equation}
It is evident from the Eq. (\ref{hor2}) that we must have the following
condition fulfilled for the existence of the BH
\begin{equation}
\left( 2M+\alpha M+\alpha
l_{0}\right) ^{2}-4\left( 1+\alpha \right) Q^{2}\geq 0.
\label{con2}
\end{equation}
Inequalities (\ref{con1}) and (\ref{con2}) together provide the parameter space
for which we have a BH. It is shown in Fig. \ref{ps}
\begin{figure}
\centering
{{\includegraphics[width=7.5cm]{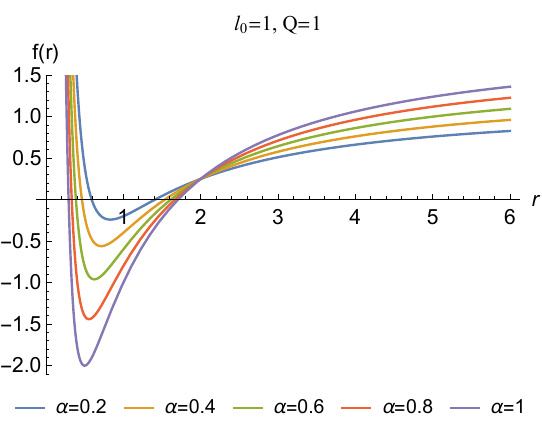} }}\qquad
{{\includegraphics[width=7.5cm]{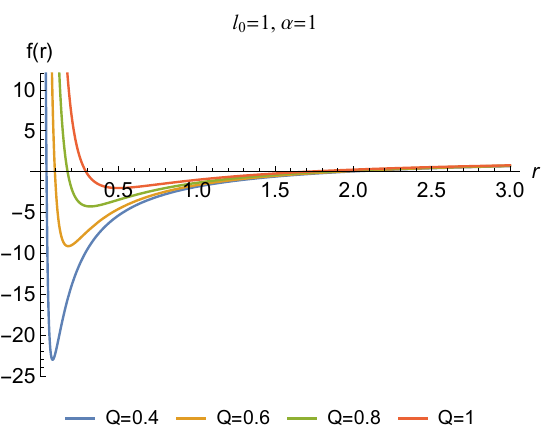}}}\qquad
{{\includegraphics[width=7.5cm]{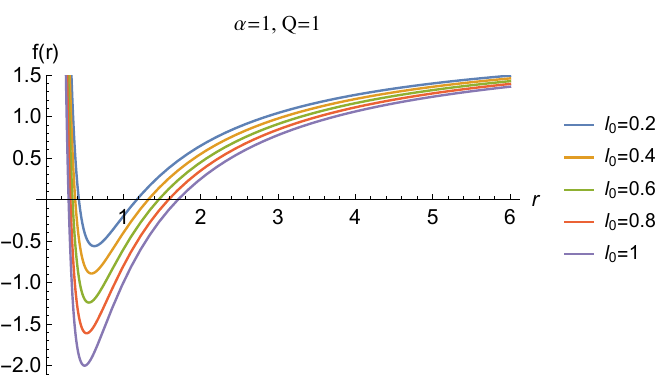}}}
\caption{The metric function (\ref{mf2}) vs. $r$ for different values of the BH parameters ($M = 1$).} \label{figa1}
\end{figure}
For a clear understanding of the parameter region enabling the BH to exist, a graph of the metric function (\ref{mf2}) is generated in  Fig. \ref{figa1}.

\section{Spinorial wave equations}

The aim of this section is to review Dirac and Rarita-Schwinger equations
in the background of the HBH solution.

\subsection{Dirac equation}

Our focus in this part is on obtaining the effective potential for fermions
propagating in HBH geometry with spin- 1/2 field. Hence, we consider the vielbein
formalism for the spin- 1/2 fields in curved spacetime (\ref{M1}), the vielbein can
be defined as follows
\begin{equation}
e_{\widehat{\alpha }}^{\mu }=daig\left( \frac{1}{\sqrt{g}},\sqrt{g},\frac{1}{%
r},\frac{1}{r\sin \theta }\right) .
\end{equation}%
The Dirac equation for massive spin-1/2 particles
\begin{equation}
\gamma ^{\mu }\left[ \left( \partial _{\mu }+\Gamma _{\mu }\right) +m\right]
\Psi =0, \label{q33}
\end{equation}%
where, $\gamma ^{\mu }$\ is the Dirac gamma matrix and $\Gamma _{\mu }$ is
the spin connection, which can be expressed in terms of the Christoffel
symbols $\Gamma _{\mu \nu }^{\rho }$ as follows%
\begin{equation}
\Gamma _{\mu }=\frac{1}{8}e_{\widehat{\alpha }}^{\rho }\left( \partial _{\mu
}e_{\rho \widehat{\beta }}-\Gamma _{\mu \rho }^{\sigma }e_{\sigma \widehat{%
\beta }}\right) \left[ \gamma ^{\widehat{\alpha }},\gamma ^{\widehat{\beta }}%
\right].
\end{equation}%
Here, the Dirac gamma matrices, $\gamma ^{\mu }$are represented in terms of
the Pauli spin matrices $\sigma ^{i}\left( i=0,1,2,3\right) .$ The solution
procedure of Dirac's equation (\ref{q33}) will be ignored in this study due to
the repetition of references \cite{Wong, badawi2023}. We will present the final
answer as a Schr\"{o}dinger-like one-dimensional wave equation with Dirac field
effective potential. Thus, the effective potentials of the massive fermionic waves
having spin- 1/2 and moving in the HBH geometry are,
\begin{equation}
V_{\pm 1/2}=\pm \frac{dW}{dr_{\ast }}+W^{2}, \label{p9}
\end{equation}%
where%
\begin{equation}
W=\frac{\left( \sqrt{g}/r\right) \sqrt{\lambda ^{2}+m^{2}r^{2}}}{1+\left(
g/2\omega \right) \left( \lambda m/\left( \lambda ^{2}+m^{2}r^{2}\right)
\right) }, \label{p10}
\end{equation}%
here, $\lambda=\left( \frac{1}{2}+s\right) $ is the standard spherical harmonics
indices, and $\omega$ is the angular frequency of the incoming field. Effective potentials
are expressed explicitly as follows:
\begin{equation}
\begin{split}
V_{\pm 1/2}=\frac{\left( \lambda^{2}+m^{2}r^{2}\right) ^{5/2}\sqrt{g}}{D^{2}}&\Bigg[
\frac{\sqrt{g}}{r^{2}}\left( \lambda^{2}+m^{2}r^{2}\right) ^{1/2}\pm \left( \frac{
g^{\prime }}{2r}-\frac{g}{r^{2}}\right) \pm 3m^{2}g \\
&\mp \frac{g}{rD}\left(
2m^{2}r+(\lambda m/2\omega )g^{\prime }\right) \Bigg], \label{p11}
\end{split}
\end{equation}
where
\begin{equation*}
D=\left( \lambda ^{2}+m^{2}r^{2}+\left( \lambda m/2\omega \right) g\right) .
\end{equation*}%
We can obtain the effective potential of massless Dirac fields (fermions)
propagating in this spacetime by setting $m=0$ in (\ref{p11}) namely
\begin{equation}
V_{\pm 1/2}=\frac{\lambda }{r^{2}}\left( \lambda g\pm \frac{r\sqrt{g} g^{\prime }}{2
}\mp g^{3/2}\right) . \label{p12}
\end{equation}
\subsection{Rarita-Schwinger equation}

We will use the massless form of the Rarita-Schwinger equation to
represent the spin- 3/2 field,
\begin{equation}
\gamma ^{\mu \nu \alpha }\widetilde{\emph{D}}_{\nu }\psi _{\alpha }=0
\label{Rs 1}
\end{equation}%
where $\widetilde{\emph{D}}_{\nu }$ is the super covariant derivative, $\psi
_{\alpha }$ indicates the spin-3/2 field and $\gamma ^{\mu \nu \alpha }$ is
the antisymmetric of Dirac gamma matrices given by
\begin{equation}
\gamma ^{\mu \nu \alpha }=\gamma ^{\lbrack \mu }\gamma ^{\nu }\gamma
^{\alpha ]}=\gamma ^{\mu }\gamma ^{\nu }\gamma ^{\alpha }-\gamma ^{\mu
}g^{\nu \alpha }+\gamma ^{\nu }g^{\mu \alpha }-\gamma ^{\alpha }g^{\mu \nu }.
\end{equation}%
The super covariant derivative for the spin-3/2 field in our BH spacetime can
be written as%
\begin{equation}
\widetilde{\emph{D}}_{\nu }=\nabla _{\nu }+\frac{1}{4}\gamma _{\alpha
}F_{\nu }^{\alpha }+\frac{i}{8}\gamma _{\nu \alpha \mu }F^{\alpha \mu }.
\end{equation}%
The solution procedure of Dirac's equation (\ref{Rs 1}) will be ignored in this
study due to the repetition of references \cite{chen16,chen18}. We will present
the final answer as a Schr\"{o}dinger-like one-dimensional wave equation with Dirac
field effective potential. Thus, the explicit forms of the effective potentials for
spin-3/2 fermions are written as
\begin{equation}
V_{1,2}=g\frac{\overline{\lambda }\left( 1-\overline{\lambda }^{2}\right) }{%
r\left( g-\overline{\lambda }^{2}\right) }\left[ \frac{\overline{\lambda }%
\left( 1-\overline{\lambda }^{2}\right) }{r\left( g-\overline{\lambda }%
^{2}\right) }\pm \left( \frac{rg^{\prime }-2g}{2r\sqrt{g}}\right) \mp \frac{%
\sqrt{g}g^{\prime }}{\left( g-\overline{\lambda }^{2}\right) }\right] ,
\label{p32}
\end{equation}
where the eigenvalue $\overline{\lambda }=\left( l+1/2\right)$ , and $
l=3/2,5/2,7/2,...$.

\section{GF\lowercase{s} of HBH}

When attempting to formulate a general description of the spectrum observed by an
observer at infinity, it is critical to know the transmission amplitude of the BH's
radiation or GFs \cite{hb20,hb21,hb22,hb23,hb24,hb25,hb26,hb27,hb28,hb29,hb30}. Our
aim of this section is to examine the GFs of HBH using the rigorous bounds
and general
semi-analytic bounds. Consequently, it is possible to determine how the potential
affects the GF. The transmission probability $\sigma _{l}\left(
w\right) $ is given by \cite{hb29,hb30}
\begin{equation}
\sigma _{l}\left( w\right) \geq \sec h^{2}\left( \int_{-\infty }^{+\infty
}\wp dr_{\ast }\right) , \label{is8}
\end{equation}
in which $r_{\ast }$ is the tortoise coordinate and
\begin{equation}
\wp =\frac{1}{2h}\sqrt{\left( \frac{dh\left( r_{\ast }\right) }{dr_{\ast }}%
\right) ^{2}+(w^{2}-V_{eff}-h^{2}\left( r_{\ast }\right) )^{2}}. \label{is9}
\end{equation}
where $h(r_{\ast })$ is a positive function satisfying $h\left( -\infty
\right) =h\left( +\infty \right) =w$. For more details, one can see \cite{hb29}. We
select $h=w$. Therefore, Eq. (\ref{is9})
\begin{equation}
\sigma _{l}\left( w\right) \geq \sec h^{2}\left( \int_{r_{h}}^{+\infty }%
\frac{V_{eff}}{2w}dr_{\ast }\right) . \label{gb1}
\end{equation}%
In this process, the metric function plays a significant role in determining
the relationship between the GFs and the effective potential. Our GF
calculations will be carried out in three cases since we have fermions with
different spins. When computing GFs, we usually focus on the
study of the potential $V_{+}$.

\subsection{Spin- 1/2 fermions emission}

In this section, we will look at the GF using rigorous bounds.
This method allows us to analyze the results qualitatively. As a result, the
potential's effect on the greybody factor can be calculated. The rigorous
bounds on the greybody factors are given by
\begin{equation}
\sigma _{l}\left( w\right) \geq \sec h^{2}\left( \int_{-\infty }^{+\infty
}\left\vert \frac{V_{eff}}{2w}\right\vert dr_{\ast }\right) \label{rb1}
\end{equation}

Substituting the effective potential (\ref{p9}) derived from Dirac equations
into Eq. (\ref{rb1}), we obtain
\begin{equation}
\sigma _{l}^{+}\left( w\right) \geq \sec h^{2}\left[ \frac{1}{2w}\left(
\int_{-\infty }^{+\infty }\left\vert \frac{dW}{dr_{\ast }}\right\vert
dr_{\ast }+\int_{-\infty }^{+\infty }\left\vert W^{2}\right\vert dr_{\ast
}\right) \right] . \label{in10}
\end{equation}%
We will discuss separately the first and second integrals in Eq. (\ref{in10}%
). For the first integral, we have%
\begin{equation}
\int_{-\infty }^{+\infty }\left\vert \pm \frac{dW}{dr_{\ast }}\right\vert
dr_{\ast }=W\left\vert _{r_{+}}^{r_{-}}=\right. 0,
\end{equation}%
where $r_{+}$ and $r_{-}$ are the two horizons (\ref{hor2}) of the BH. The
second integral can be written as
\begin{equation}
\int_{r_{+}}^{r-}\left( \frac{\left( \lambda ^{2}+m^{2}r^{2}\right) ^{2}}{%
r^{2}\left[ (\lambda m/2k)\left\vert g\right\vert +\left( \lambda
^{2}+m^{2}r^{2}\right) \right] }\right) dr, \label{in11}
\end{equation}%
There is a considerable difference between the results of this formulation
for massless and massive instances. These two cases are, therefore, considered
separately.

\subsubsection{Massless case}

\begin{figure}
\centering
{{\includegraphics[width=7.5cm]{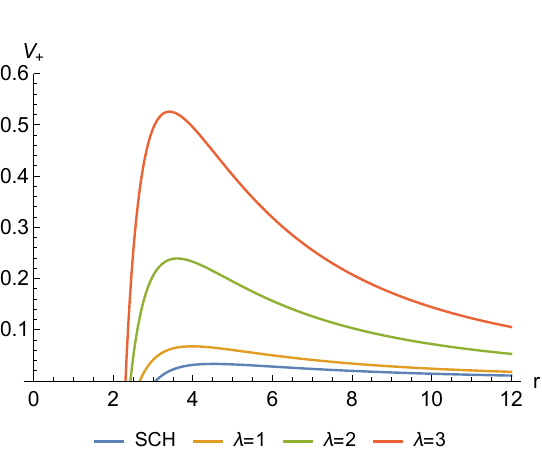} }}\qquad
{{\includegraphics[width=7.5cm]{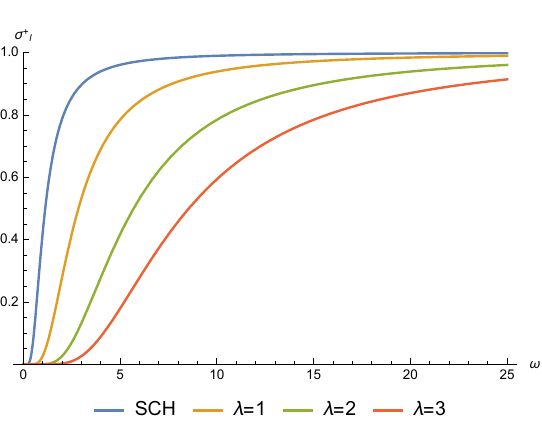}}}
\caption{The left panel shows the potential (\ref{p12}) for massless
spin- $1/2$ field with $M = 1$, $\alpha=0.5$, $l_{0}=1$ and $Q=0.7$. The right
panel shows the corresponding
GF bound.} \label{fig1}
\end{figure}

\begin{figure}
\centering
{{\includegraphics[width=7.5cm]{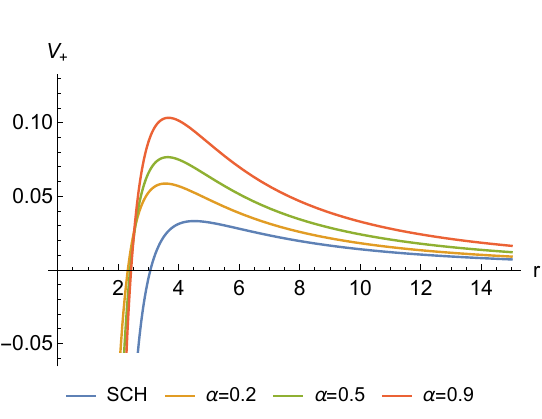} }}\qquad
{{\includegraphics[width=7.5cm]{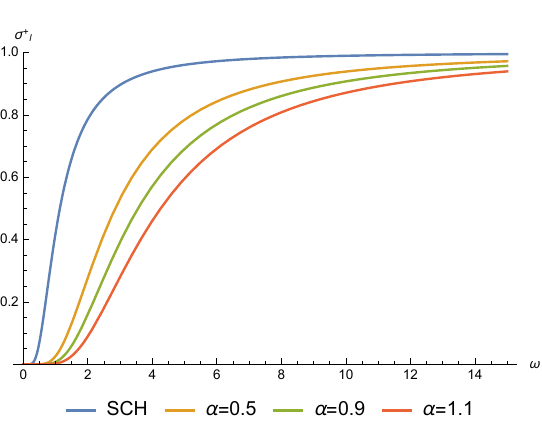}}}
\caption{The left panel shows the potential (\ref{p12}) for massless
spin- $1/2$ field with $M = 1$, $\lambda=1$, $l_{0}=1$ and $Q=0.6$. The right
panel shows the corresponding
GF bound.} \label{fig2}
\end{figure}

\begin{figure}
\centering
{{\includegraphics[width=7.5cm]{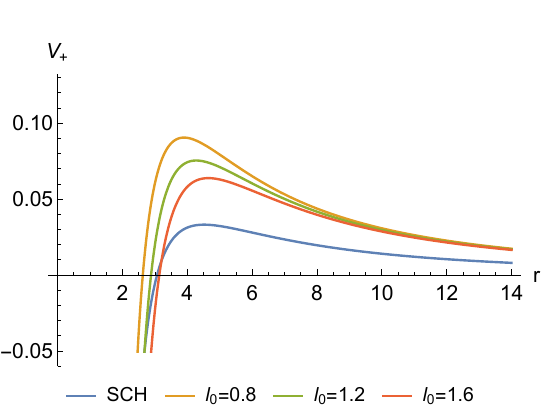} }}\qquad
{{\includegraphics[width=7.5cm]{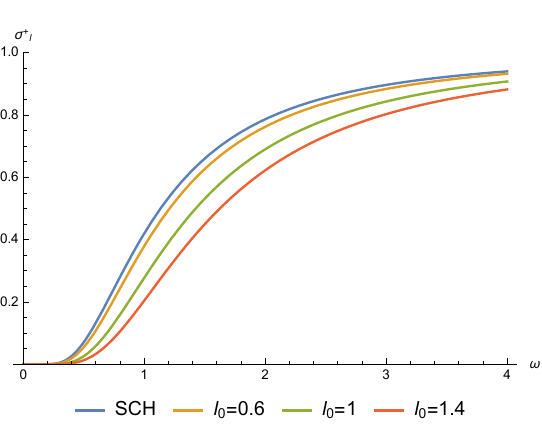}}}
\caption{The left panel shows the potential (\ref{p12}) for massless
spin- $1/2$ field with M = 1, $\lambda=1$, $\alpha=0.8$ and $Q=0.4$. The right
panel shows the corresponding
GF bound.} \label{fig3}
\end{figure}

\begin{figure}
\centering
{{\includegraphics[width=7.5cm]{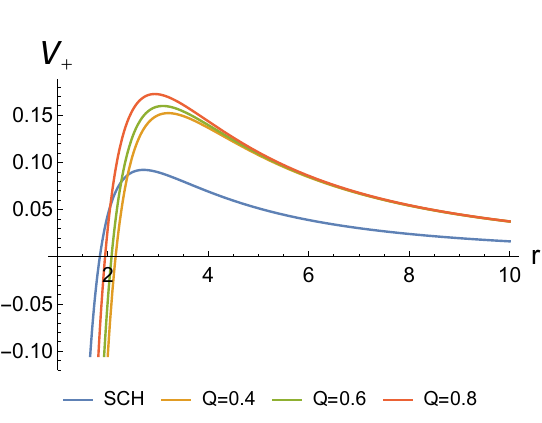} }}\qquad
{{\includegraphics[width=7.5cm]{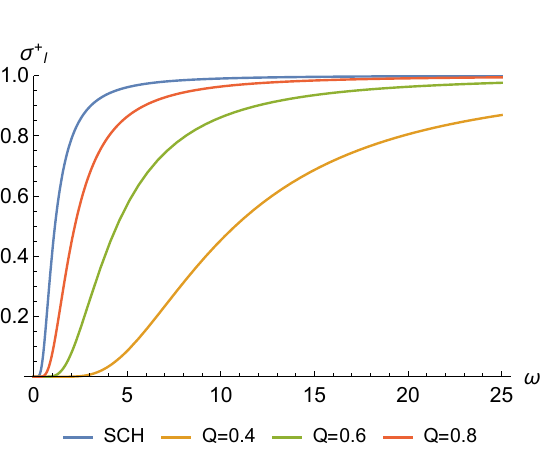}}}
\caption{The left panel shows the potential (\ref{p12}) for massless
spin- $1/2$ field with $M = 1$, $\lambda=1$, $\alpha=1$ and $l_{0}=1.4$.
The right panel shows the corresponding
GF bound.} \label{fig4}
\end{figure}

To compute the GFs for massless spin- 1/2 fermions emission, we choose $m=0,$
then the integral (\ref{in11}) becomes%
\begin{equation}
\int_{r_{+}}^{r-}\frac{\lambda ^{2}}{r^{2}}dr=\lambda ^{2}\left( \frac{1}{%
r_{+}}-\frac{1}{r_{-}}\right) .
\end{equation}%
Substituting the result of this integral in Eq. (\ref{in10}), the rigorous
bound can be expressed as%
\begin{equation}
\sigma _{l}^{+}\left( w\right) \geq \sec h^{2}\left( \frac{\lambda ^{2}}{%
2\omega }\left[ \frac{1}{r_{+}}-\frac{1}{r_{-}}\right] \right) . \label{in1}
\end{equation}%
After putting the values of the horizons (\ref{hor2}), the rigorous
bound of the HBH for massless fermions is calculated as follows:%
\begin{equation}
\sigma _{l}^{+}\left( w\right) \geq \sec h^{2}\left( \frac{\lambda ^{2}}{%
2\omega }\left[ \frac{-\sqrt{\left( 2M+\alpha M+\alpha l_{0}\right)
^{2}-4\left( 1+\alpha \right) Q^{2}}}{Q^{2}}\right] \right) . \label{gf12}
\end{equation}%
A massless spin- $1/2$ fermion bound in an HBH exhibits a behavior that depends on
the distance between two horizons, as reflected in the argument of function $%
sech$. It is, therefore, possible to analyze the behavior of the bound by
considering how the distance between two horizons varies with the hairy
parameters. These can be illustrated in Figs. \ref{fig1}, \ref{fig2}, \ref{fig3} and \ref{fig4}.
Using the potential shape as a starting point, we can examine how the rigorous
bound $\sigma _{l}$ behaves. This can be achieved by varying the hairy
parameters, $( \alpha, Q,l_{0})$ and angular parameter $\lambda$. By adjusting
the hairy parameters, the potential increases when $\lambda$ increases, as shown
in the left panel of Fig. \ref{fig1}. As shown in the right panel of Fig. \ref{fig1},
the GF decreases for a given value of $\omega$ because the wave is more difficult
to transmit through the higher potential. Similar analysis can be performed for
the hairy parameters $(\alpha$ and $l_0)$, as shown in Figs. \ref{fig2} and
\ref{fig3}. The analysis for the hairy parameter $Q$, on the other hand, is
the inverse, as shown in Fig. \ref{fig4}.

\subsubsection{Massive case}

To compute the GFs for massive spin- 1/2 fermions emission, we will write
the integral (\ref{in11}) as%
\begin{equation}
\int_{r_{+}}^{r-}\frac{\lambda ^{2}\left( 1+\mu ^{2}r^{2}\right) }{%
r^{2}\left( 1+\frac{\mu g}{\left( 1+\mu ^{2}r^{2}\right) 2w}\right) }%
dr=\int_{r_{+}}^{r-}Adr, \label{in33}
\end{equation}%
where
\begin{equation}
A=\frac{\lambda ^{2}\left( 1+\mu ^{2}r^{2}\right) }{r^{2}\left( 1+\frac{\mu g%
}{\left( 1+\mu ^{2}r^{2}\right) 2w}\right) },\hspace{1cm}\mu =m/\lambda .
\end{equation}%
When we consider the equation above, we can see that $A$ is larger than $1$,
since the factor $1+\frac{\mu g}{\left( 1+\mu ^{2}r^{2}\right) 2w}>1$. In
this way, we can approximate the integrand, which is given by the following
\begin{equation}
A=\frac{\lambda ^{2}\left( 1+\mu ^{2}r^{2}\right) }{r^{2}\left( 1+\frac{\mu g%
}{\left( 1+\mu ^{2}r^{2}\right) 2w}\right) }\leq \frac{\lambda ^{2}}{r^{2}}%
\left( 1+\mu ^{2}r^{2}\right) =A_{app}.
\end{equation}%
\begin{figure}
\centering
{{\includegraphics[width=7.5cm]{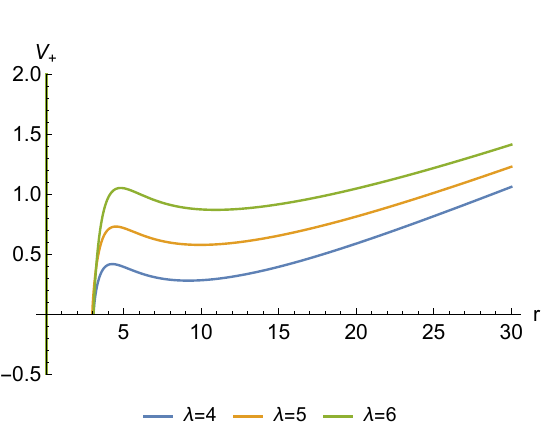} }}\qquad
{{\includegraphics[width=7.5cm]{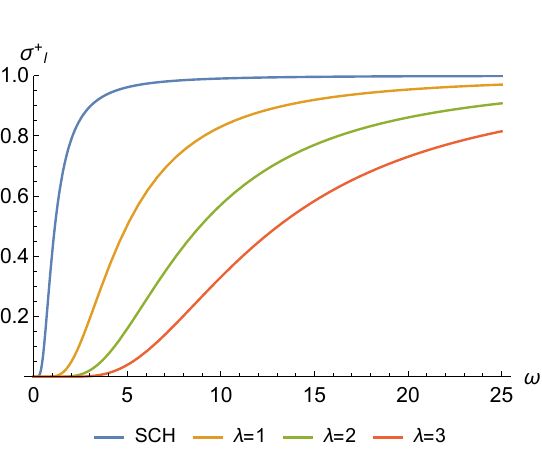}}}
\caption{The left panel shows the potential (\ref{p11}) for massive spin- 1/2 field with  $M = 1$, $m=0.5$, $Q=0.7$ and $l_{0}=0.5$. The right panel shows the corresponding GF bound} \label{fig5}
\end{figure}
\begin{figure}
\centering
{{\includegraphics[width=7.5cm]{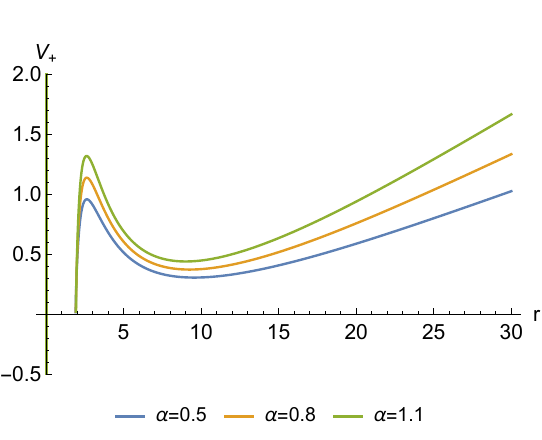} }}\qquad
{{\includegraphics[width=7.5cm]{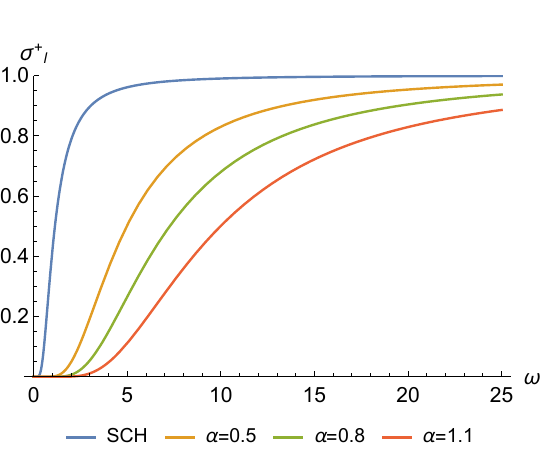}}}
\caption {The left panel shows the potential (\ref{p11}) for massive spin- 1/2 field for varying $Q$ with  $M = 1$, $m=1$, $\alpha=0.4$ and $l_{0}=1$. The right panel shows the corresponding GF bound} \label{figa5}
\end{figure}

\begin{figure}
\centering
{{\includegraphics[width=7.5cm]{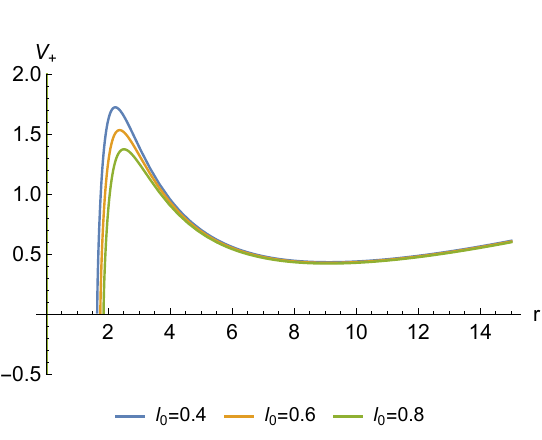} }}\qquad
{{\includegraphics[width=7.5cm]{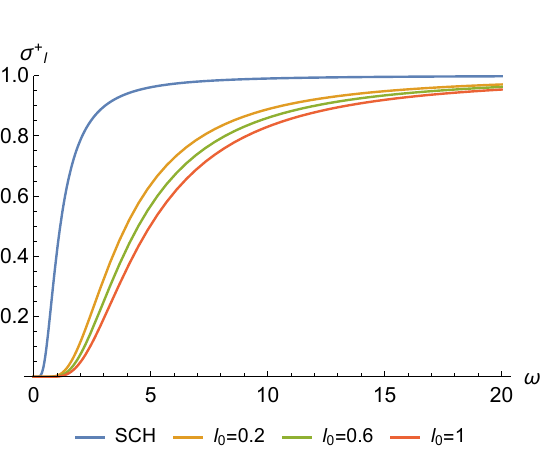}}}
\caption{The left panel shows the potential (\ref{p11}) for massive spin- 1/2 field with  $M = 1$, $m=1$, $Q=0.4$ and $\alpha=1$. The right panel shows the GFs of massive spin- $1/2$ field for varying $l_{0}$} \label{fig6}
\end{figure}
\begin{figure}
\centering
{{\includegraphics[width=7.5cm]{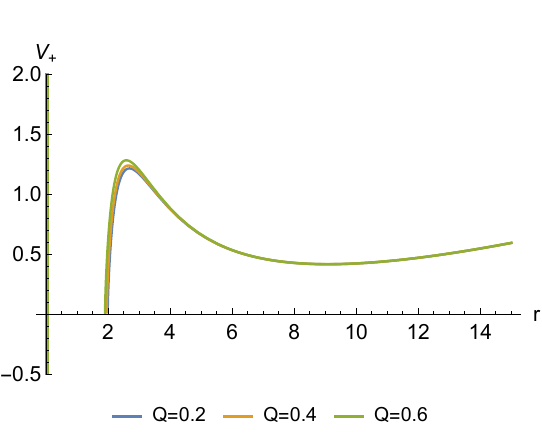} }}\qquad
{{\includegraphics[width=7.5cm]{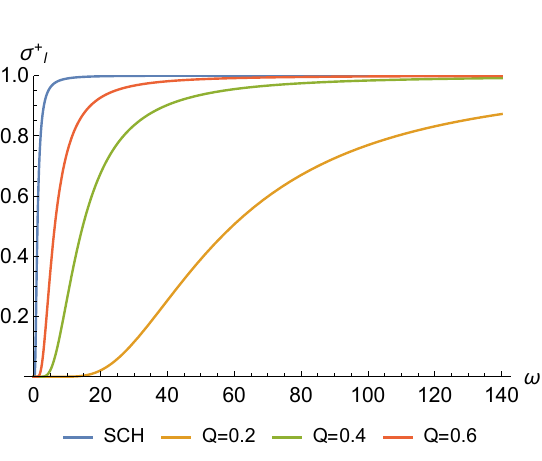}}}
\caption{The left panel shows the potential (\ref{p11}) for massive spin- 1/2 field with  $M = 1$, $m=1$, $l_{0}=1$ and $\alpha=1$. The right panel shows the GFs of massive spin- $1/2$ field for varying $Q$} \label{figa6}
\end{figure}
The integral Eq. (\ref{in33}) can be evaluated by using the same arguments
as in \cite{hb33} namely,
\begin{equation}
\sigma _{l}^{+}\left( w\right) \geq \sec h^{2}\left(
\int_{r_{+}}^{r-}A_{app}dr\right)
\end{equation}%
\begin{equation}
\sigma _{l}^{+}\left( w\right) \geq \sec h^{2}\left( \frac{\lambda ^{2}}{%
2\omega }\left[ \frac{r_{-}-r_{+}}{r_{+}r_{-}}\left( 1+\mu
^{2}r_{+}r_{-}\right) \right] \right) . \label{gf13}
\end{equation}%
After putting the values of the horizons in Eq. (\ref{gf13}), the rigorous
bound of the HBH for massive fermions is calculated as follows
\begin{equation}
\sigma _{l}^{+}\left( w\right) \geq \sec h^{2}\left( \frac{\lambda ^{2}}{%
2\omega }\left[ \frac{-\sqrt{\left( 2M+\alpha M+\alpha l_{0}\right)
^{2}-4\left( 1+\alpha \right) Q^{2}}}{\left( 1+\alpha \right) Q^{2}}\left(
1+\alpha +\mu ^{2}Q^{2}\right) \right] \right) .\label{boud1}
\end{equation}%
This expression is reduced to that of the GF bound for the massless case when
$\mu \rightarrow 0.$ According to (\ref{boud1}), the bound for the massive case
is still reliant on the hairy parameters in the same way that the bound for the
massless case is. The GF bounds for massive spin- $1/2$ are illustrated in
Figs. \ref{fig5}- \ref{figa6}. It is worth to mention that according to Ref \cite{olliv}, HBHs were created by demanding that they satisfy the strong energy condition or dominant energy condition between the BH's horizon $r\geq 2M$ and infinity. As shown in \cite{olliv}, all of the new HBHs solutions correspond to Schwarzschild vacuum deformations. Therefore, the plots of the effective potential exhibits a negative gap when the $r$ coordinate is less than the horizon as shown in the left panel of the Figs. \ref{fig5}- \ref{figa6}. 

\subsection{Spin- 3/2 fermions emission}

\begin{figure}
\centering
{{\includegraphics[width=7.5cm]{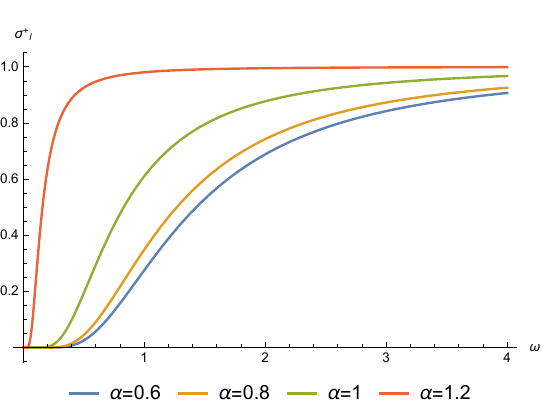} }}\qquad
{{\includegraphics[width=7.5cm]{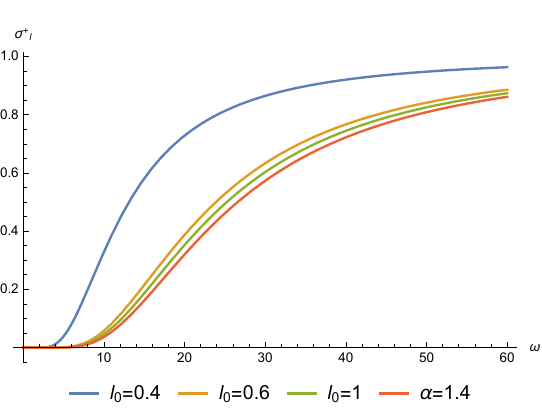}}}
\caption{The GFs of massless spin- $3/2$ field for varying $\alpha$ (left panel) and
$l_{0}$ (right panel) with $M = 1$, $m=0.5$, $Q=0.7$ and $l_{0}=0.5$.} \label{fig7}%
\end{figure}

With general semi-analytic bounds, we can obtain the GF for spin-3/2 fermions
emission using the potential derived in (\ref{p32}). Then, Eq. (\ref{gb1}) becomes
\begin{equation}
\begin{split}
\sigma _{l}\left( w\right) \geq \sec h^{2}&\Bigg[
\frac{\lambda }{2\omega }
\int_{r_{h}}^{+\infty }\Bigg( \frac{\overline{\lambda }\left( 1-\overline{
\lambda }^{2}\right) }{r^{2}\left( g-\overline{\lambda }^{2}\right) ^{2}} \\
& \left( \overline{\lambda }\left( 1-\overline{\lambda }%
^{2}\right) +\frac{\left( rg^{\prime }-2g\right) \left( g-\overline{\lambda }%
^{2}\right) }{2\sqrt{g}}-r\sqrt{g}g\right) dr \Bigg) \Bigg], \label{in18}
\end{split}
\end{equation}
We will apply asymptotic series expansion to overcome the difficulties encountered
in evaluating the above complicated integral (\ref{in18}). Then, the GF of massless
spin- $3/2$ fermions is
\begin{equation}
\begin{split}
\sigma _{l}^{+}\left( w\right) \geq \sec h^{2}&\Bigg[
\frac{1}{2w}\Bigg(
\frac{\overline{\lambda }\left( 1-\overline{\lambda }^{2}\right) \left(
2\alpha +\alpha ^{2}+\left( 2\overline{\lambda }\sqrt{1+\alpha }-1-\overline{
\lambda }^{2}\right) \right) \left( 2M+\alpha \left( l_{0}+M\right) \right)
}{2\overline{\lambda }\sqrt{1+\alpha }\left( 1+\alpha -\overline{\lambda }
^{2}\right) r_{h}^{2}} \\
& +\frac{\overline{\lambda }\left( 1-\overline{\lambda }^{2}\right)
\left( \alpha \sqrt{1+\alpha }-\left( \sqrt{1+\alpha }-\overline{\lambda }
\right) \left( \overline{\lambda }^{2}-1\right) \right) }{\left( 1+\alpha -
\overline{\lambda }^{2}\right) r_{h}}\Bigg) \Bigg],
\end{split}
\end{equation}
Figure \ref{fig7} shows the variation of the GF of massless spin-$3/2$ field with various
hairy parameters. The graph shows that as $\alpha$ parameter increases, GF decreases,
whereas GF increases as $l_{0}$ increases.

\section{Ringdown waveform}

With the help of the time domain integration method \cite{gundlach1}, we, in this
section, intend to scrutinize the impact of parameters $(\alpha$,$ Q, l_0)$ on the
time evolution of the massless fermionic spin -1/2 and spin -3/2 fields. We implement
the time domain integration method with initial conditions
$\psi(r_*,t) = \exp \left[ -\dfrac{(r_*-\hat{r}_{*})^2}{2\sigma^2} \right]$
and $\psi(r_*,t)\vert_{t<0} = 0$. We choose the values of $\Delta t$ and
$\Delta r_{*}$ such that the Von Neumann stability condition, $\frac{\Delta t}{\Delta r_*} < 1$,
is satisfied. The time profile helps us scrutinize the qualitative variation of the decay
rate and the frequency with the variation of parameters $\alpha$,$ Q, l_0$.

\begin{figure}[H]
\centering
\subfigure[]{
\label{tp1}
\includegraphics[width=7.5cm]{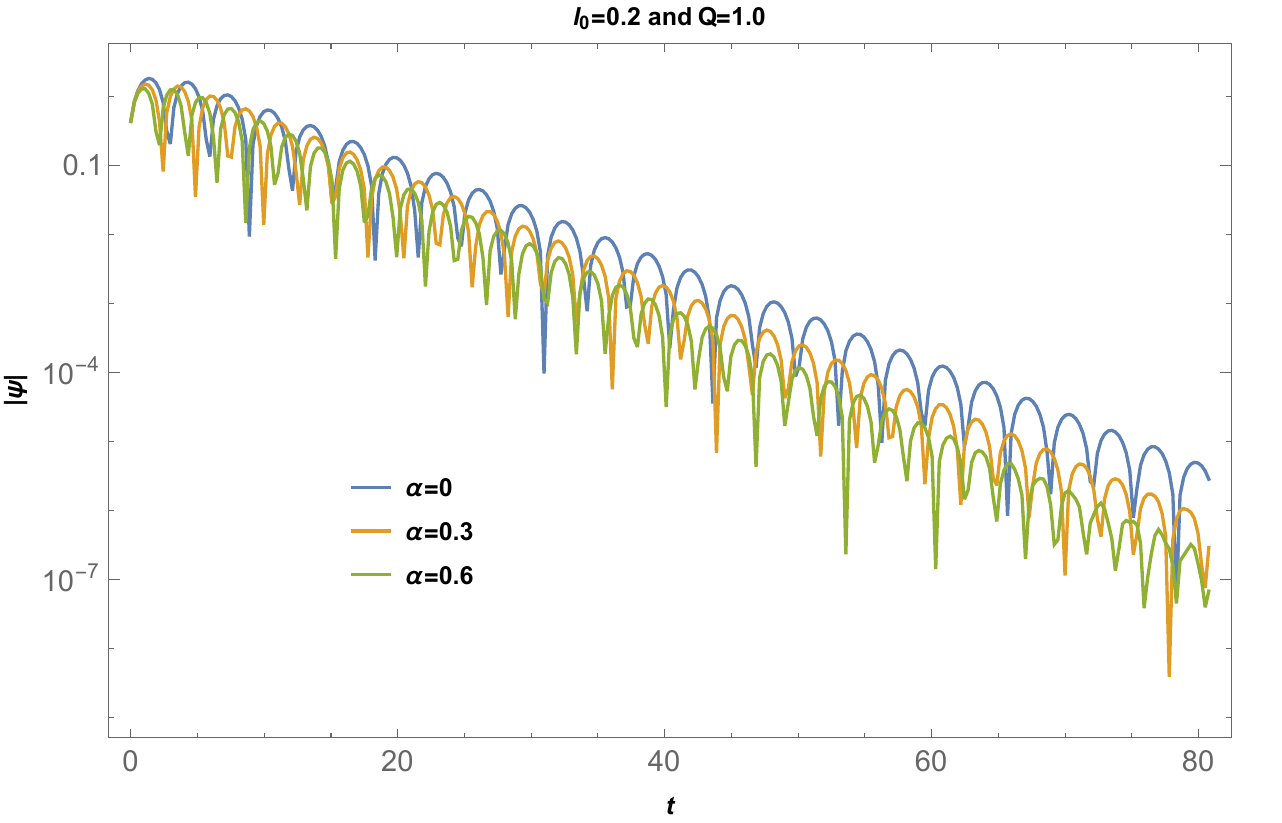}}
\subfigure[]{
\label{tp2}
\includegraphics[width=7.5cm]{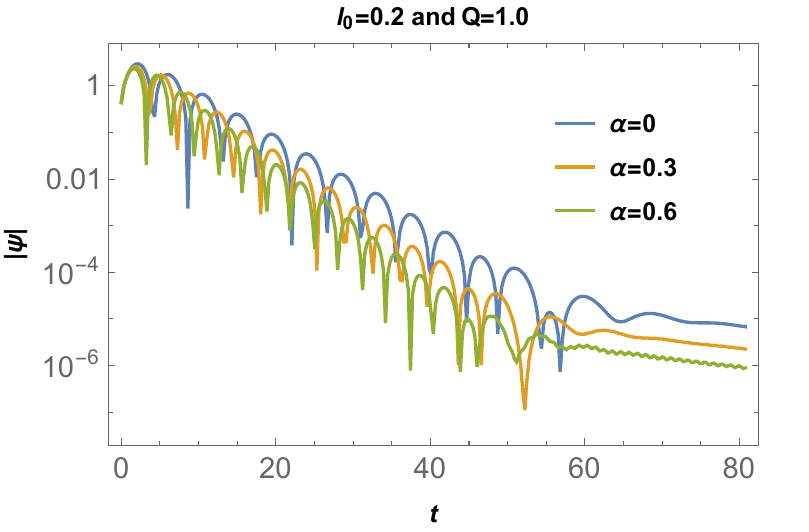}
}
\subfigure[]{
\label{tp3}
\includegraphics[width=7.5cm]{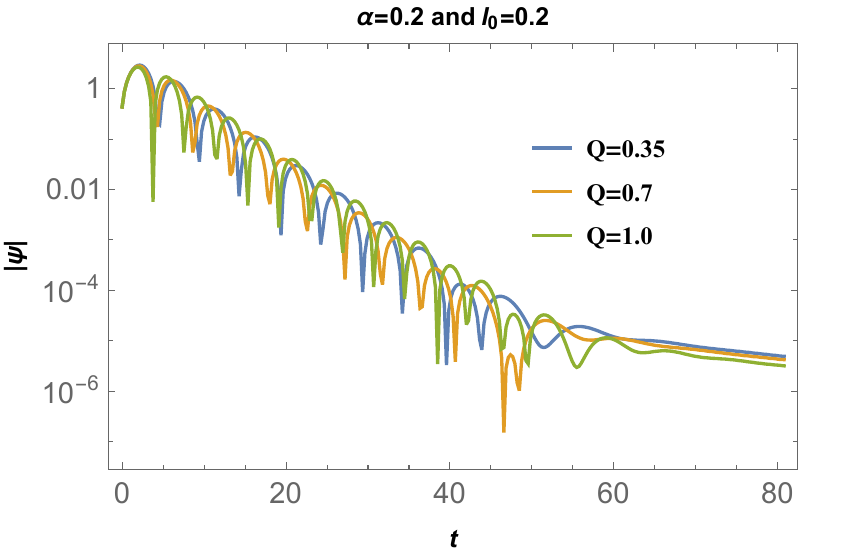}
}
\caption{Time profile of massless fermionic spin -1/2 perturbation. Here, we have taken $\lambda=3$.}
\label{tp}
\end{figure}
\begin{figure}[H]
\centering
\subfigure[]{
\label{tp13}
\includegraphics[width=7.5cm]{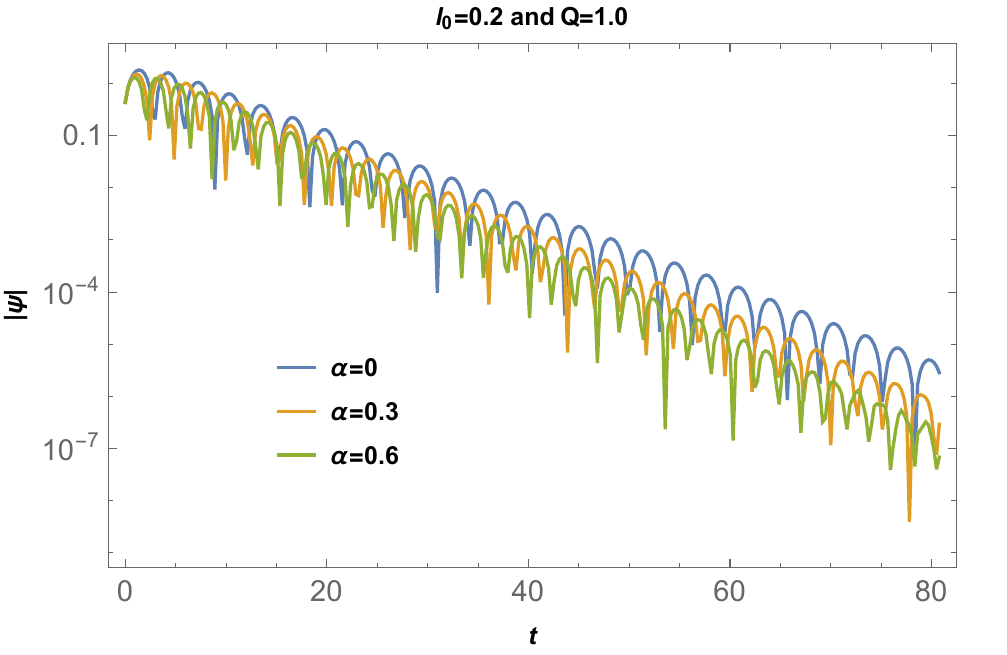}
}
\subfigure[]{
\label{tp23}
\includegraphics[width=7.5cm]{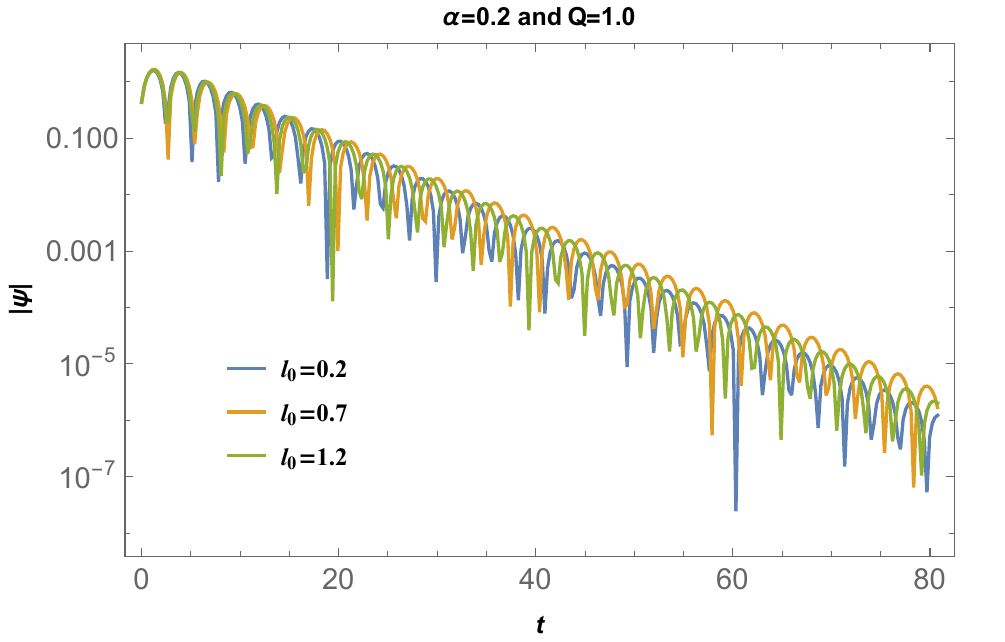}
}
\subfigure[]{
\label{tp33}
\includegraphics[width=7.5cm]{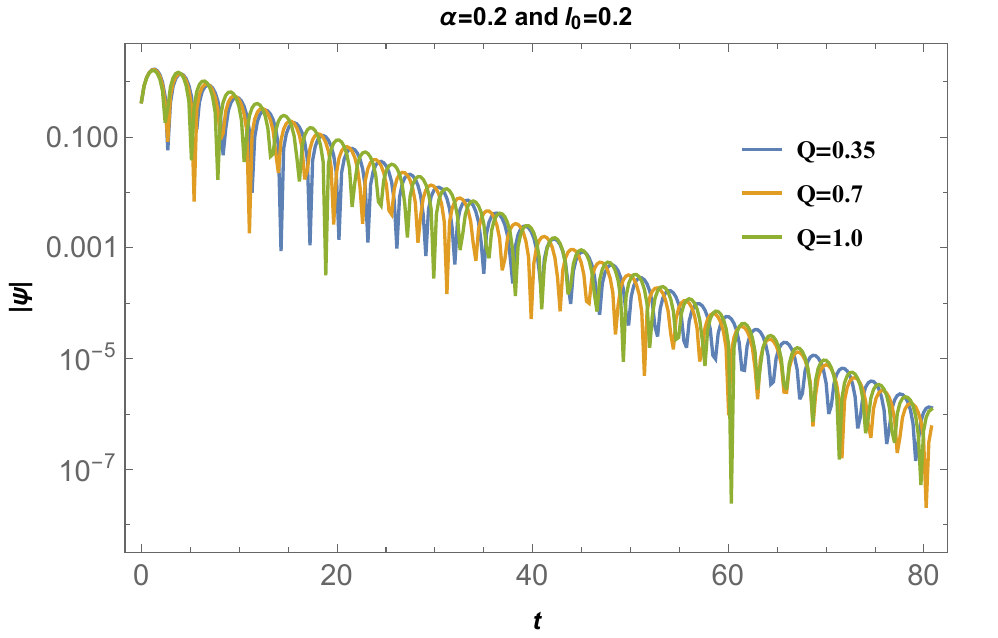}
}
\caption{Time profile of massless fermionic spin- 3/2 perturbation. Here, we have
taken $\overline{\lambda}=3$.}
\label{tp3f}
\end{figure}
\begin{figure}[H]
\centering
\includegraphics[width=7.5cm]{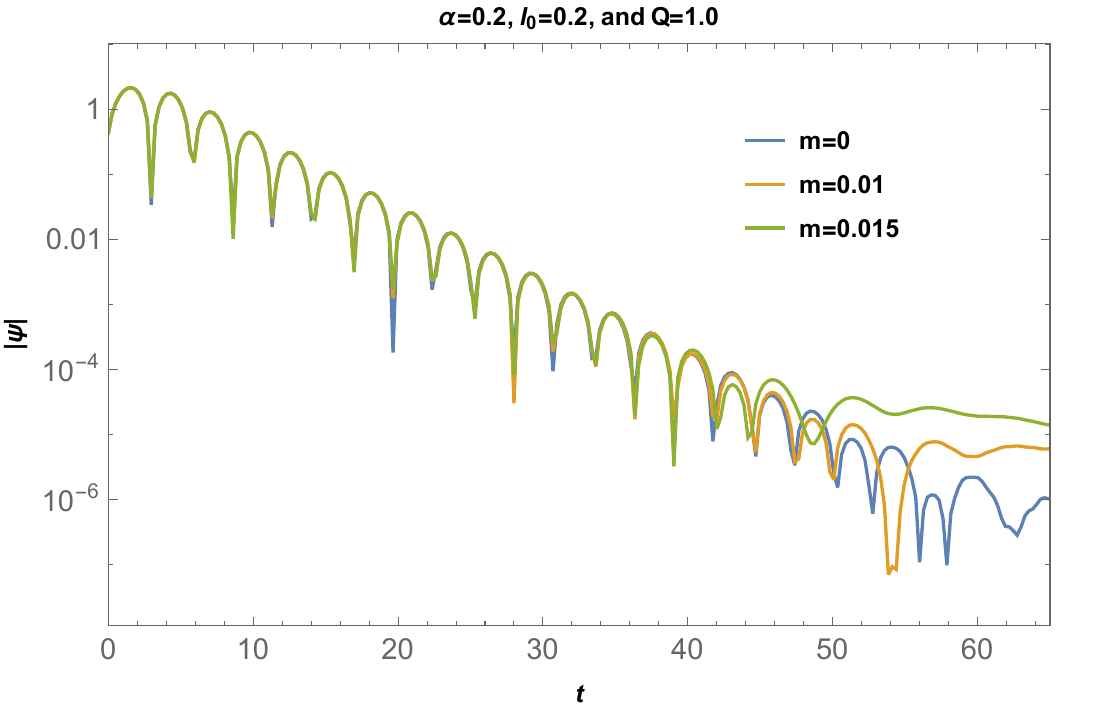}
\caption{Time profile of massless and massive fermionic spin- 1/2 perturbations. Here, we have
taken $\lambda=2$.}
\label{compare}
\end{figure}

In Fig. \ref{tp}, we show the ringdown waveform for a massless spin -$1/2$ perturbation
field, and in Fig. \ref{tp3f}, we provide the waveform for a massless spin -$3/2$ field.
From Figs \ref{tp1}, \ref{tp13}, we observe that the decay rate as well the frequency
increases with an increase in $\alpha$ for both the perturbations. The parameter $l_0$
influences quasinormal modes in such a way that the decay rate, as well as the frequency, first decreases and then increases with an increase in $l_0$. The impact of $Q$ can be
inferred from Figs. \ref{tp3}, \ref{tp33}. We can conclude that the decay rate decreases
and the frequency increases with $Q$.
To compare the ringdown waveforms of massless and massive spin -1/2 perturbations, we plot them for massless and massive perturbations in Fig. \ref{compare}. It clearly shows that the frequency and decay rate for massive perturbation is less than those for massless perturbation. 
Our study in this section conclusively illustrates the impact of parameters $(\alpha, Q, l_0)$
on the time evolution of perturbation fields and enriches our understanding of this very important aspect of BH physics.

\section{Hawking spectra and sparsity}

\begin{figure}[H]
\centering
\subfigure[]{
\label{pl1}
\includegraphics[width=7.5cm]{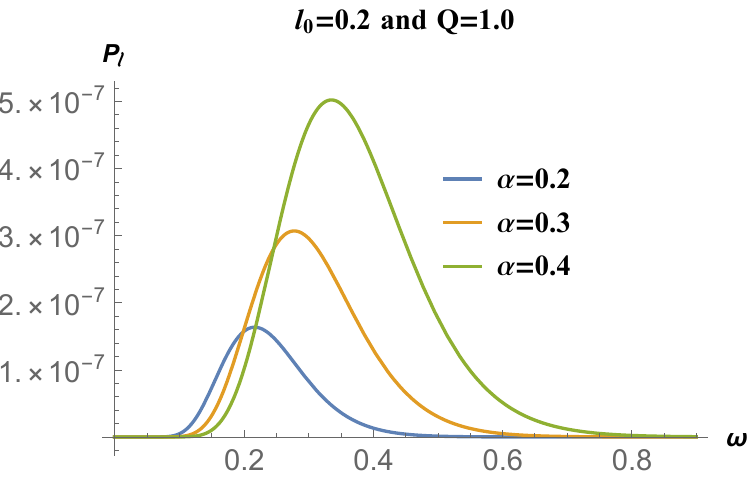}
}
\subfigure[]{
\label{pl2}
\includegraphics[width=7.5cm]{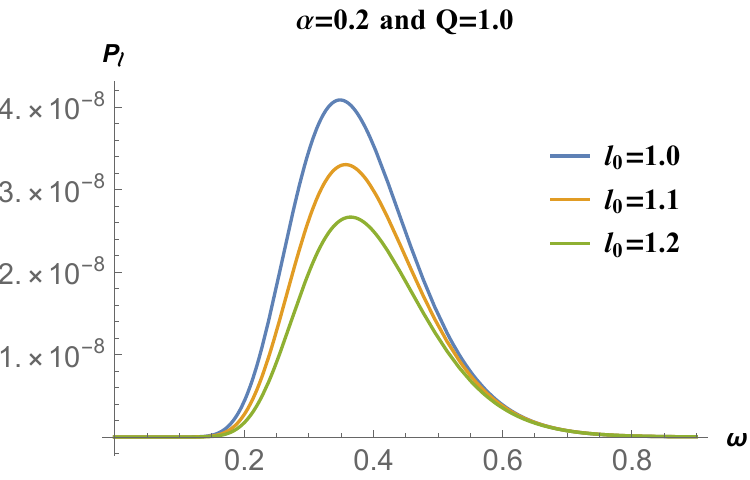}
}
\subfigure[]{
\label{pl3}
\includegraphics[width=7.5cm]{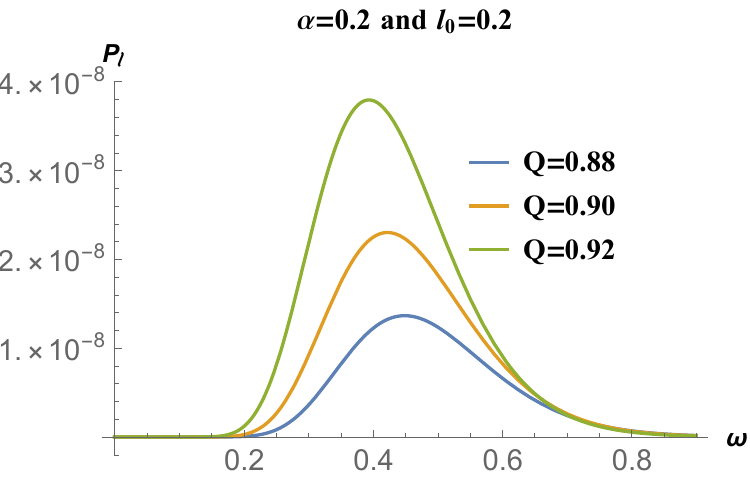}
}
\caption{Power spectrum of massive fermionic spin -1/2 field. Here, we have taken
$\lambda=2$ and $\mu=0.25$.}
\label{plfig}
\end{figure}

\begin{figure}[H]
\centering
\subfigure[]{
\label{pl13}
\includegraphics[width=7.5cm]{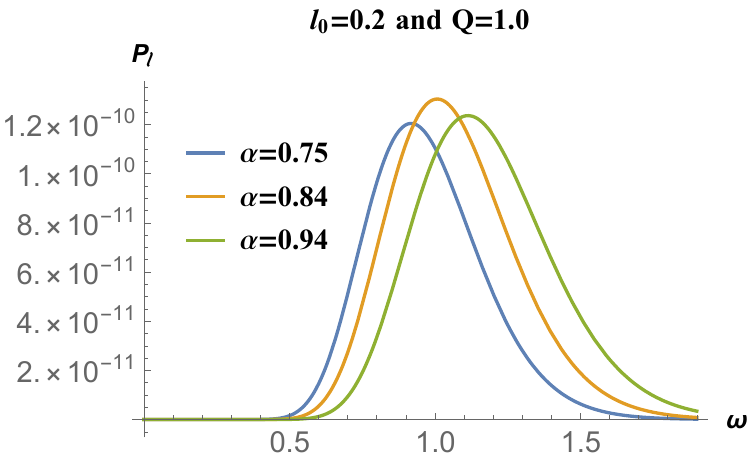}
}
\subfigure[]{
\label{pl23}
\includegraphics[width=7.5cm]{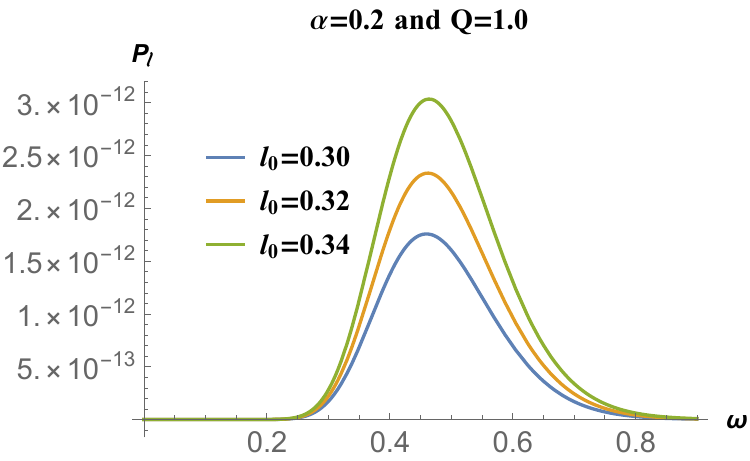}
}
\subfigure[]{
\label{pl33}
\includegraphics[width=7.5cm]{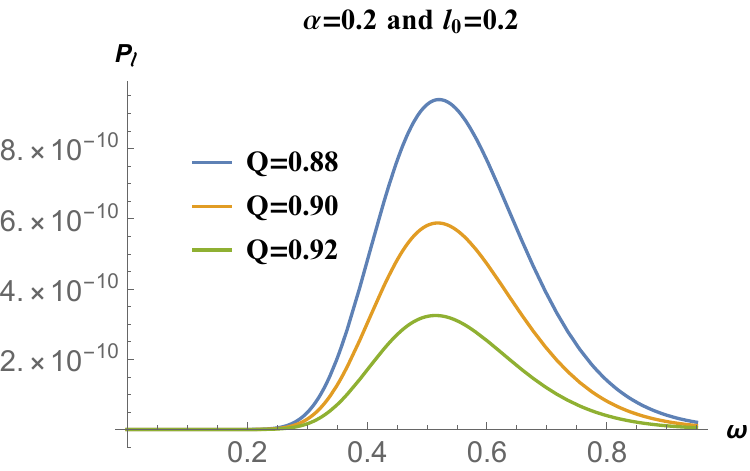}
}
\caption{Power spectrum of massless fermionic spin -3/2 field. Here, we have
taken $\overline{\lambda}=2$.}
\label{plfig3f}
\end{figure}

The Hawking spectra and the sparsity of radiation emitted by HBH are investigated in this
section. The power emitted per unit frequency in the $l^{th}$ mode is \cite{yg2017, fg2016}
\begin{equation}\label{pl}
P_l\left(\omega\right)=\frac{A}{8\pi^2}\sigma_{l}^{+}\frac{\omega^3}{e^{\omega/T_{H}}-1},
\end{equation}
where A is taken to be the horizon area \cite{yg2017}. $T_H$ is the Hawking temperature given by
\begin{eqnarray}\label{th}
T_{H}&=&\frac{1}{4\pi} \frac{dg}{dr}|_{r=r_{+}}\\\nonumber
&=&\frac{H1}{H2} \quad \text{where}\\\nonumber
H1&=&(\alpha +1)^2 \left(\alpha ^2 \left(l_0+M\right){}^2+2 M \left(\sqrt{\left(\alpha \left(l_0+M\right)+2 M\right){}^2-4 (\alpha +1) Q^2}+2 M\right)\right.\\\nonumber
&&\left. +\alpha\left(l_0+M\right) \left(\sqrt{\left(\alpha \left(l_0+M\right)+2 M\right){}^2-4 (\alpha +1) Q^2}+4 M\right)-4 \alpha Q^2-4 Q^2\right),\\\nonumber
H2&=&\pi \left(\alpha \left(l_0+M\right)+\sqrt{\left(\alpha l_0+(\alpha +2) M\right){}^2-4 (\alpha +1) Q^2}+2 M\right){}^3.
\end{eqnarray}
The total power emitted in the $l^{th}$ mode is given by
\begin{equation}
P_{tot}= \int_{0}^{\infty} P_{\l}\left(\omega \right) d \omega.\label{ptot}
\end{equation}
Combining Eqs. (\ref{pl}, \ref{th}) we analyze the qualitative variation of $P_l$ with
respect to $\omega$ for different scenarios in Figs. \ref{plfig}, \ref{plfig3f}. From
Fig. \ref{pl1} we observe that the total power emitted by the BH increases with $\alpha$, and
the position of the peak shifts towards the right. Figs. \ref{pl2}, \ref{pl3} also reveal
that the total emitted power decreases with $l_0$ but increases with $Q$, whereas the peak
of the spectrum shifts towards the right for an increase in $l_0$ and towards the left for
an increase in $Q$. These observations are for a massive fermionic spin -$1/2$ field. Now,
for a massless spin -$3/2$ field, we turn our attention towards Fig. \ref{plfig3f}. From Fig.
\ref{pl13}, we observe that the peak of the power spectrum first increases with $\alpha$ and
then decreases, whereas the position of the peak always shifts to the right. It is also
observed that the total power emitted by the BH increases with $\alpha$. It is evident
from Figs. \ref{pl23}, \ref{pl33} that the total power emitted by the BH increases with
$l_0$ and decreases with $Q$.

To have a quantitative measure of the radiation emitted by the BH, we introduce
sparsity, a dimensionless quantity, defined as \cite{yg2017, fg2016, ac2020, sh2016, sh2015}
\begin{equation}
\eta=\frac{\tau_{gap}}{\tau_{emission}}.
\label{eta}
\end{equation}
Here, $\tau_{gap}$ and $\tau_{emission}$ are the time that is taken by a radiation
quantum for emissions and the average time gap between two successive radiation quanta,
respectively. These are defined by
\begin{equation}\label{tgap}
\tau_{gap}=\frac{\omega_{max}}{P_{tot}} \quad \tau_{emission} \geq \tau_{localisation}=\frac{2 \pi}{\omega_{max}},.
\end{equation}
where $\tau_{localisation}$ is the time period of the emitted wave of frequency $\omega_{max}$.
For $\eta\ll1$, we will have a continuous flow of Hawking radiation, whereas a large value
of $\eta$ implies a sparse Hawking radiation. Quantitative values of $\omega_{max}$, $P_{max}$, $P_{tot}$,
and $\eta$ are given in Tables \ref{alpha2}, \ref{l02}, \ref{Q2} for a massive spin -$1/2$ field and
in Tables \ref{alpha3}, \ref{l03}, \ref{Q3} for a massless spin -$3/2$ field. These tables reinforce
our conclusions drawn from Figs.\ref{plfig}, \ref{plfig3f}. Additionally, we can also observe the
effect of various parameters on the sparsity. From Tables \ref{alpha2}, \ref{l02}, \ref{Q2} we see
that the sparsity of the radiation decreases with $\alpha$ and $Q$ but increases with $l_0$ for a
massive spin -$1/2$ field. It can be observed from the Table \ref{alpha3} that the sparsity
initially decreases and then increases with $\alpha$ having critical value at $\alpha=0.871$. On the other hand, $\eta$ decreases with
$l_0$ and increases with $Q$ as evident from Tables \ref{l03}, \ref{Q3}.
\begin{table}[!htp]
\centering
\caption{Numerical values of $\omega_{max}$, $P_{max}$, $P_{tot}$, and $\eta$ for massive
spin -1/2 perturbation for various values of $\alpha$ with $l_0=0.2$, $Q=1$, $\lambda=2$,
and $\mu=0.25$.}
\setlength{\tabcolsep}{0pt}
\begin{tabular}{|c|c|c|c|c|c|}
\hline
$\alpha$ & 0.2 & 0.4 & 0.6 & 0.8 & 1.0 \\
\hline
$\omega _{\max }$ & 0.217541 & 0.335618 & 0.445538 & 0.552871 & 0.656921 \\
\hline
$P_{max}$ & $1.63372$ $\times$ $10^{-7}$ & $5.0177$ $\times$ $10^{-7}$ & $1.0799$ $\times$ $10^{-6} $ & $1.96344$ $\times$ $10^{-6}$ & $3.21797$ $\times$ $10^{-6}$ \\
\hline
$ P_{\text{tot}}$ &$ 2.58744\times 10^{-8}$ & $1.22075\times 10^{-7}$ &$ 3.47708\times 10^{-7}$ & $7.82162\times 10^{-7}$ & $1.524\times 10^{-6}$ \\
\hline
$ \eta$ & $2.91094\times 10^5 $& $1.46853\times 10^5$ &$ 9.08604\times 10^4$ & $6.21972\times 10^4$ & $4.50672\times 10^4$ \\
\hline
\end{tabular}
\label{alpha2}
\end{table}

\begin{table}[!htp]
\centering
\caption{Numerical values of $\omega_{max}$, $P_{max}$, $P_{tot}$, and $\eta$ for massive
spin -1/2 perturbation for various values of $l_0$ with $\alpha=0.2$, $Q=1$, $\lambda=2$ and $\mu=0.25$.}
\setlength{\tabcolsep}{0pt}
\begin{tabular}{|c|c|c|c|c|c|}
\hline
$l_0 $ & $0.2$ & $0.4$ & $0.6$ & $0.8$ & $1.0$ \\
\hline
$ \omega _{\max }$ & $0.217541$ & $0.268992$ & $0.304746$ & $0.331781$ & $0.359994$ \\
\hline
$ P_{max }$ &$ 1.63372\times 10^{-7}$ & $1.32654\times 10^{-7}$ & $9.29835\times 10^{-8}$ & $6.21794\times 10^{-8}$ & $4.05768\times 10^{-8}$ \\
\hline
$ P_{\text{tot}}$ &$ 2.58744\times 10^{-8}$ & $2.51919\times 10^{-8}$ & $1.94188\times 10^{-8}$ & $1.37758\times 10^{-8}$ & $9.41305\times 10^{-9}$ \\
\hline
$\eta$ & $2.91094\times 10^5$ & $4.57127\times 10^5$ & $7.61159\times 10^5$ & $1.27176\times 10^6$ & $2.19119\times 10^6$ \\
\hline
\end{tabular}
\label{l02}
\end{table}

\begin{table}[!htp]
\centering
\caption{Numerical values of $\omega_{max}$, $P_{max}$, $P_{tot}$, and $\eta$ for massive
spin -1/2 perturbation for various values of $Q$ with $\alpha=0.2$, $l_0=0.2$, $\lambda=2$ and $\mu=0.25$.}
\setlength{\tabcolsep}{0pt}
\begin{tabular}{|c|c|c|c|c|c|}
\hline
$Q $ & $0.35$ & $0.48$ & $0.61$ & $0.74$ & $0.87 $ \\
\hline
$ \omega _{\max }$ & $1.38934$ & $0.9631$ & $0.862479$ & $0.599048$ & $0.436239$ \\
\hline
$P_{\max }$ & $1.65373\times 10^{-22}$ & $2.06223\times 10^{-16}$ & $6.96079\times 10^{-13}$ & $1.84607\times 10^{-10}$ & $1.0156\times 10^{-8}$ \\
\hline
$P_{\text{tot}}$ & $8.17937\times 10^{-23}$ & $9.30337\times 10^{-17}$ & $2.95924\times 10^{-13}$ & $6.21224\times 10^{-11}$ & $2.96887\times 10^{-9}$ \\
\hline
$ \eta $ & $3.75593\times 10^{21}$ & $1.5868\times 10^{15}$ & $4.00071\times 10^{11}$ & $9.19382\times 10^8$ & $1.02018\times 10^7$ \\
\hline
\end{tabular}
\label{Q2}
\end{table}

\begin{table}[!htp]
\centering
\caption{Numerical values of $\omega_{max}$, $P_{max}$, $P_{tot}$, and $\eta$ for massless
spin -3/2 perturbation for various values of $\alpha$ with $l_0=0.2$, $Q=1$, $\overline{\lambda}=2$.}
\setlength{\tabcolsep}{0pt}
\begin{tabular}{|c|c|c|c|c|c|}
\hline
$\alpha $ & $0.2$ & $0.4$ & $0.6$ & $0.8$ & $1.0 $ \\
\hline
$ \omega _{\max }$ & $0.436239$ & $0.599048$ & $0.862479$ & $0.9631$ & $1.12591$ \\
\hline
$P_{\max }$ & $2.75106\times 10^{-13}$ & $1.56319\times 10^{-11}$ & $6.72487\times 10^{-11}$ & $1.27798\times 10^{-10}$ & $1.09201\times 10^{-10}$ \\
\hline
$ P_{\text{tot}}$ &$ 5.93889\times 10^{-14}$ & $4.91794\times 10^{-12}$ & $3.09834\times 10^{-11}$ & $6.46398\times 10^{-11}$ & $6.82237\times 10^{-11}$ \\
\hline
$\eta$ & $5.09993\times 10^{11}$ & $1.16134\times 10^{10}$ & $3.82109\times 10^9$ & $2.28383\times 10^9 $& $2.95727\times 10^9$ \\
\hline
\end{tabular}
\label{alpha3}
\end{table}

\begin{table}[!htp]
\centering
\caption{Numerical values of $\omega_{max}$, $P_{max}$, $P_{tot}$, and $\eta$ for massless
spin -3/2 perturbation for various values of $l_0$ with $\alpha=0.2$, $Q=1$, $\overline{\lambda}=2$.}
\setlength{\tabcolsep}{0pt}
\begin{tabular}{|c|c|c|c|c|c|}
\hline
$l_0$ & $0.2$ & $0.4$ & $0.6$ & $0.8$ & $1.0$ \\
\hline
$\omega _{\max }$ & $0.436239$ & $0.436239$ & $0.436239$ & $0.436239$ &$ 0.436239$ \\
\hline
$P_{\max }$ & $2.75106\times 10^{-13}$ & $5.64809\times 10^{-12}$ & $2.68204\times 10^{-11}$ & $7.27206\times 10^{-11}$ & $1.47412\times 10^{-10}$ \\
\hline
$ P_{\text{tot}}$ & $5.93889 \times 10^{-14}$ & $1.45152\times 10^{-12}$ & $7.46322\times 10^{-12}$ & $2.07892\times 10^{-11}$ & $4.21614\times 10^{-11}$ \\
\hline
$ \eta $ & $5.09993\times 10^{11}$ & $2.08664\times 10^{10}$ & $4.0583\times 10^9$ & $1.45691\times 10^9$ & $7.18382\times 10^8$ \\
\hline
\end{tabular}
\label{l03}
\end{table}

\begin{table}[!htp]
\centering
\caption{Numerical values of $\omega_{max}$, $P_{max}$, $P_{tot}$, and $\eta$ for massless
spin -3/2 perturbation for various values of $Q$ with $\alpha=0.2$, $l_0=0.2$, $\overline{\lambda}=2$.}
\setlength{\tabcolsep}{0.2mm}
\begin{tabular}{|c|c|c|c|c|c|}
\hline
$Q $& $0.35$ & $0.48$ & $0.61$ & $0.74$ & $0.87$ \\
\hline
$\omega _{\max }$ & $0.536861$ & $0.536861$ & $0.536861$ &$ 0.536861$ & $0.536861$ \\
\hline
$P_{\max }$ & $1.86991\times 10^{-8}$ & $1.50416\times 10^{-8}$ & $1.03636\times 10^{-8}$ &$ 5.23738\times 10^{-9}$ & $1.13655\times 10^{-9}$ \\
\hline
$ P_{\text{tot}}$ & $6.08023\times 10^{-9}$ & $4.85163\times 10^{-9}$ & $3.30778\times 10^{-9}$ & $1.64503\times 10^{-9}$ & $3.43338\times 10^{-10} $\\
\hline
$ \eta $ & $7.54438\times 10^6$ & $9.45489\times 10^6$ & $1.38678\times 10^7$ & $2.78849\times 10^7$ & $1.33605\times 10^8$ \\
\hline
\end{tabular}
\label{Q3}
\end{table}

\section{Conclusion}
We begin the assessment of how the rigorous bound $\sigma _{l}$
behaves, comprehending the potential shape of the potential
associated with fulminic perturbation. The shape of the potential
depends crucially on the hairy parameters, $( \alpha, Q, l_{0}$), as
well as on the angular parameter $\lambda$. According to the
findings, the left panel of Fig. \ref{fig1} displays that the potential rises with the
increase in the parameter $\lambda$ for particular hairy parameters. The right
panel of Fig. \ref{fig1}, however, shows that the GF decreases with
$\lambda$ for a given value of $\omega$ because the wave is more
difficult to transmit through the higher potential. Similar
analysis follows for the hairy parameters $(\alpha$ and $l_0)$ as
it is transparent from the Figs. \ref{fig2} and \ref{fig3}. Fig.
\ref{fig4} displays that the action of the hairy parameter
$Q$, on the other hand, is entirely the inverse of what was
found for the parameter $(\alpha$ and $l_0)$.

Using the argument \cite{hb33} leads us to find out the GFs
bounds for both massive and massless cases. As per (\ref{boud1}),
the massive case's bound depends on the hairy
parameters, just as the bound for the massless case. The GFs
bounds for massive spin- $1/2$ fermion are depicted in Figs.
 \ref{fig5}- \ref{figa6}. Along with the spin- $1/2$ fermion, we
have considered the spin-$3/2$ fermion in our study, where
Dirac equation is replaced by the Schwinger-Dyson equation. Fig.
(\ref{fig7}) shows the variation of the GF of massless spin-$3/2$
fermion with various hairy parameters. The graph indicates that
while GF increases as $l_{0}$ increases, it drops as $\alpha$
parameter increases.

Our study includes the ringdown waveform due to the perturbation
of massless spin -$1/2$ and spin -$3/2$ fields. Fig. \ref{tp} displays
the ringdown waveform for a massless spin -$1/2$
The perturbation field in which Fig. \ref{tp3f} displays the
same due to the perturbation of spin -$3/2$ field. From Figs.
(\ref{tp1}, \ref{tp13}) we observe that the decay rate, as well as the
frequency, increases with an increase in $\alpha$ for both the
perturbations caused by both of the fields having the odd
integral spin $1/2$ and $3/2$. The influences of the parameter
$l_0$ on the quasinormal modes are as follows. Both the decay rate
and the frequency of oscillation decrease, to begin with,
and then increase with an increase in $l_0$. The impact of $Q$
can be inferred from Figs. (\ref{tp3}, \ref{tp33}). It is
found that the decay rate decreases, whereas the frequency
increases when the value of $Q$ enhances.

When the study of the emissive power of Hawking radiation is carried
out, we analyze the qualitative variation of $P_l$ with respect to
$\omega$ for different scenarios in Figs. \ref{plfig},
\ref{plfig3f}. Combining Eqs. (\ref{pl}, \ref{th}) we became
able to study the qualitative behavior of the variation of $P_l$
with respect to $\omega$. According to Fig. \ref{pl1}, the overall power emission rises with $\alpha$, and the
location of the peak of the spectra moves to the right. Figs.
(\ref{pl2}, \ref{pl3}) reveal that the total emitted power
decreases with $l_0$ but increases with $Q$, whereas the peak of
the spectra acquire a right shift and left shift, respectively, for
an increase in $l_0$ and $Q$. These findings pertain to the
perturbation when it is caused due to the massive fermionic field
having spin -$1/2$. For the perturbation due to the massless
spin -$3/2$ fermionic field, let us focus on Fig.
\ref{plfig3f}. Fig. \ref{pl13} shows that the peak of the power
spectrum increases with $\alpha$, to begin with, and then is followed
by a drop, whereas the position of the peak of the spectra always
shifts to the right. It is also observed that the total power
emitted by the BH increases with $\alpha$. The Figs. \ref{pl23},
and \ref{pl33} exhibit that the total power emitted by the BH
enhances with $l_0$; however, it drops off with Q.

For a quantitative idea, we compute the numerical values of
$\omega_{max}$, $P_{max}$, $P_{tot}$, and $\eta$ which are
displayed in Tables \ref{alpha2}, \ref{l02}, and \ref{Q2} for a
massive spin -$1/2$ fermionic field, and for a massless spin
-$3/2$ field, the corresponding quantities are presented in Tables
\ref{alpha3}, \ref{l03}, and \ref{Q3}. These tables reinforce our
conclusions drawn from Figs.\ref{plfig}, and \ref{plfig3f}.
Additionally, we can observe the effect of various parameters on
the sparsity of the Hawking radiation. From Tables \ref{alpha2},
\ref{l02}, and \ref{Q2}, we see that the sparsity of the radiation
expands with $\alpha$ and $Q$ but shrinks with $l_0$ for a massive
spin -$1/2$ field. Table \ref{alpha3} furnishes that the sparsity
initially diminishes and then enhances with $\alpha$. Conversely, $\eta$ falls off with $l_0$ and grows with $Q$ according
to the numerical information of the Tables \ref{l03} and
\ref{Q3}. According to our study, hairy parameters have significant effects on GFs, QNMs, Hawking spectra, and sparsity. These findings will enhance our understanding of hairy black holes and their astrophysical significance.
\\
{\LARGE Acknowledgements}\newline
We are thankful to the Editor and anonymous Referees for their constructive suggestions and comments.\\

\bigskip
{\Large Data Availability}
\newline No data availability in this manuscript.


\begin{thebibliography}{the}
\bibitem{HAW1} S. W. Hawking, Commun. Math. Phys. 25 (1972) 15215
\bibitem{HAW2} S. W. Hawking, M. J. Perry and A. Strominger, Phys. Rev. Lett. 116 (2016) 231301 [arXiv:1601.00921 [hep-th]].
\bibitem{PTS} T. P. Sotiriou and V. Faraoni, Phys. Rev. Lett. 108 (2012) 081103 [arXiv:1109.6324 [gr-qc]].
\bibitem{BABI} E. Babichev and C. Charmousis, JHEP 1408 (2014) 106 [arXiv:1312.3204 [gr-qc]].
\bibitem{PTS1} T. P. Sotiriou and S. Y. Zhou, Phys. Rev. Lett. 112 (2014) 251102 [arXiv:1312.3622 [gr-qc]].
\bibitem{RADU} C. A. R. Herdeiro and E. Radu, Int. J. Mod. Phys. D 24 (2015) 1542014 [arXiv:1504.08209 [gr-qc]].
\bibitem{SALGA} P. Canate, L. G. Jaime and M. Salgado, Class. Quant. Grav. 33 (2016) 155005 [arXiv:1509.01664 [gr-qc]].
\bibitem{RBENK} R. Benkel, T. P. Sotiriou and H. Witek, Class. Quant. Grav. 34 (2017) no.6, 064001 [arXiv:1610.09168 [gr-qc]].
\bibitem{ANTO} G. Antoniou, A. Bakopoulos and P. Kanti, "Evasion of No-Hair Theorems in Gauss-Bonnet Theories," arXiv:1711.03390 [hep-th].
\bibitem{ANAB} A. Anabalon, A. Cisterna and J. Oliva, Phys. Rev. D 89, 084050 (2014) [arXiv:1312.3597 [gr-qc]].
\bibitem{ACIST1} A. Cisterna and C. Erices, Phys. Rev. D 89, 084038 (2014) [arXiv:1401.4479 [gr-qc]].
\bibitem{ACISR2} A. Cisterna, M. Hassaine, J. Oliva and M. Rinaldi, Phys. Rev. D 96, no. 12, 124033 (2017) [arXiv:1708.07194 [hep-th]].
\bibitem{SVOLKOV} M. S. Volkov and D. V. Galtsov, JETP Lett. 50 (1989) 346 [Pisma Zh. Eksp. Teor. Fiz. 50 (1989) 312].
\bibitem{KAN1} P. Kanti, N. E. Mavromatos, J. Rizos, K. Tamvakis and E. Winstanley, Phys. Rev. D 54 (1996) 5049 [hep-th/9511071].
\bibitem{KAN2} P. Kanti, N. E. Mavromatos, J. Rizos, K. Tamvakis and E. Winstanley, Phys. Rev. D 57 (1998) 6255 [hep-th/9703192].
\bibitem{VOLKOV} M. S. Volkov and D. V. Galtsov, Phys. Rept. 319 (1999) 1 [hep-th/9810070].
\bibitem{CMART} C. Martinez, R. Troncoso, J. Zanelli, Phys.Rev. D 70 (2004) 084035 arXiv:hep-th/0406111
\bibitem{KGZ} K. G. Zloshchastiev, Phys. Rev. Lett. 94 (2005) 121101 [hep-th/0408163].
\bibitem{olliv} J. Ovalle, R. Casadio, E. Contreras, and A. Sotomayor, Phys. Dark Univ. 31, 100744 (2021),
arXiv:2006.06735 [gr-qc]
\bibitem{PTS3} T. P. Sotiriou, Class. Quant. Grav. 32 (2015) 214002 [arXiv:1505.00248 [gr-qc]].
\bibitem{OVALLE0} J. Ovalle, Phys. Lett. B788, 213 (2019), arXiv:1812.03000 [gr-qc].
\bibitem{OVALLE00} J. Ovalle, Phys. Rev. D95, 104019 (2017), arXiv:1704.05899 [gr-qc].
\bibitem{tt1} H. T. Cho, A. S. Cornell, Jason Doukas, and Wade Naylor Phys. Rev. D 75, 104005 (2007).
\bibitem{ahmad1} Ahmad Al-Badawi and Amani Kraishan, Annals of Physics 458 (2023) 169467.
\bibitem{ahmad2} Ahmad Al-Badawi and Amani Kraishan, Chinese Journal of Physics,  87  59–69 (2024).

\bibitem{OVALLE1} J. Ovalle, Mod. Phys. Lett. A23, 3247 (2008), arXiv:grqc/0703095 [gr-qc].
\bibitem{OVALLE2} J. Ovalle and R. Casadio, Buy Beyond Einstein Gravity: The Minimal Geometric
Deformation Approach in the Brane-World, Springer, 2020
\bibitem{ALTG} R. Konoplya and A. Zhidenko, Phys. Lett. B 756, 350 (2016).
\bibitem{MERGE1} R. Konoplya and A. Zhidenko, Phys. Lett. B 756, 350 (2016) [arXiv:1602.04738 [gr-qc]]; JCAP 1612, no. 12, 043 (2016) [arXiv:1606.00517 [gr-qc]];
\bibitem{MERGE2}N. Yunes, K. Yagi and F. Pretorius, Phys. Rev. D 94, no. 8, 084002 (2016) [arXiv:1603.08955 [gr-qc]].
\bibitem{MERGE4} R. A. Konoplya and A. Zhidenko, Rev. Mod. Phys. 83, 793 (2011) [arXiv:1102.4014 [gr-qc]];
\bibitem{MERGE5} E. Berti, V. Cardoso and A. O. Starinets, Class. Quant. Grav. 26, 163001 (2009)
[arXiv:0905.2975 [gr-qc]];
\bibitem{MERGE6} K. D. Kokkotas and B. G. Schmidt, Living Rev. Rel. 2, 2 (1999) [gr-qc/9909058]
\bibitem{LIGO1}B. P. Abbott et al. (LIGO Scientific and Virgo Collaborations), Phys. Rev. Lett. 116, 061102 (2016).
\bibitem{LIGO2}B. P. Abbott et al. (LIGO Scientific and Virgo Collaborations), Phys. Rev. Lett. 116, 221101 (2016).
\bibitem{QNM1} R. A. Konoplya and A. Zhidenko, Rev. Mod. Phys. 83, 793 (2011) [arXiv:1102.4014 [gr-qc]];
\bibitem{QNM2} E. Berti, V. Cardoso and A. O. Starinets, Class. Quant. Grav. 26, 163001 (2009) [arXiv:0905.2975 [gr-qc]];
\bibitem{QNM3} K. D. Kokkotas and B. G. Schmidt, Living Rev. Rel. 2, 2 (1999) [gr-qc/9909058]
\bibitem{PRESS} H. W. Press, Longwave trains of gravitational waves
from a vibrating black hole, Astrophys. J 170, L105-L108 (1971)
\bibitem{VISH} V. C. Vishveshwara, Scattering of gravitational radiation by
a Schwarzschild black hole, Nature. 227:936-938 (1970)
\bibitem{BERTI} E. Berti, V. Cardoso, A. O. Starinets, Class. Quant. Grav. 26, 163001 (2009).
\bibitem{KONOZ}R. A. Konoplya, A. Zhidenko, Rev. Mod. Phys. 83, 793 (2011).
\bibitem{AHAR} O. Aharony, S. S. Gubser, J. Maldacena, H. Ooguri, Y. Oz, Phys. Rept. 323, 183 (2000).
\bibitem{HORO} G. T. Horowitz, V. E. Hubeny, Phys. Rev. D 62, 024027 (2000).
\bibitem{SON} D. T. Son, A. O. Starinets, JHEP 09, 042 (2002).
\bibitem{ STAR} A. O. Starinets, Phys. Rev. D 66, 124013 (2002).
\bibitem{NUN} A. Nunez, A. O. Starinets, Phys. Rev. D 67, 124013 (2003).
\bibitem{KOV} P. K. Kovtun, A. O. Starinets, Phys. Rev. D 72, 086009 (2005).
\bibitem{SHOD} S. Hod, Phys. Rev. Lett. 81, 4293 (1998).
\bibitem{OD} O. Dreyer, Phys. Rev. Lett. 90, 081301 (2003).
\bibitem{MAGGI} M. Maggiore, Phys. Rev. Lett. 100, 141301 (2008).
\bibitem{KONO} R. A. Konoplya, Phys. Rev. D 70, 047503 (2004).
\bibitem{KIEF} C. Kiefer, Class. Quant. Grav. 21, L123 (2004).
\bibitem{QGP1}P. Kovtun, D. T. Son, and A. O. Starinets, Phys. Rev. Lett. 94, 111601 (2005).
\bibitem{QGP2}M. Luzum and P. Romatschke, Phys. Rev. C 78, 034915
(2008); 79, 039903(E) (2009).

\bibitem{hb10} N. Chatzifotis, C. Vlachos, K. Destounis, and
E. Papantonopoulos, Gen. Rel. Grav. 54, 49 (2022),
arXiv:2109.02678 [gr-qc].

\bibitem{hb11} Yi Yang, Dong Liu, Ali Övgün, Zheng-Wen Long, and Zhaoyi Xu, Phys. Rev. D 107, 064042 (2023).

\bibitem{hb12}R. T. Cavalcanti, R. C. de Paiva, R. da Rocha, Eur. Phys. J. Plus (2022) 137:1185.

\bibitem{hb1} J. Ovalle, Internat. J. Modern Phys. D 18 (2009) 837–852, arXiv:0809.3547.

\bibitem{hb12b} Subhash Mahapatra and Indrani Banerjee, Physics of the Dark Universe, 39
101172, (2023).

\bibitem{hb2} J. Ovalle, Modern Phys. Lett. A 25 (2010) 3323–3334, arXiv:1009.3674.

\bibitem{hb3} R. Casadio, J. Ovalle, Phys. Lett. B715 (2012) 251–255, arXiv:1201.6145.

\bibitem{hb4} J. Ovalle, F. Linares, A. Pasqua, A. Sotomayor, Classical Quantum Gravity
30 (2013) 175019, arXiv:1304.5995.

\bibitem{hb5} J. Ovalle, L.A. Gergely, R. Casadio, Classical Quantum Gravity 32 (2015) 045015, arXiv:1405.0252.

\bibitem{hb6} R. Casadio, J. Ovalle, R. da Rocha, Europhys. Lett. 110 (4) (2015) 40003,
arXiv:1503.02316.

\bibitem{hb7} R. Casadio, J. Ovalle, R. da Rocha, Classical Quantum Gravity 32 (21)
(2015) 215020, arXiv:1503.02873.

\bibitem{hb8} R.T. Cavalcanti, A.G. da Silva, R. da Rocha, Classical Quantum Gravity 33
(21) (2016) 215007, arXiv:1605.01271.

\bibitem{Wong}{R. N. P. Wongjun, C.H. Chen, Physical Review D 101, 124033 (2020).}

\bibitem{badawi2023}{A. Al-Badawi, The European Physical Journal C 83, 380 (2023).}

\bibitem{chen16} C.-H. Chen, H. T. Cho, A. S. Cornell, and G. Harmsen, Phys. Rev. D 94, 044052 (2016).

\bibitem{chen18} C.-H. Chen, H. T. Cho, A. S. Cornell, G. Harmsen, and X. Ngcobo, Phys. Rev. D 97, 024038
(2018).


\bibitem{hb20} M. Visser, Phys. Rev. A 59, 427 (1999).

\bibitem{hb21} P. Boonserm and M. Visser, Phys. Rev. D 78, 101502 (2008).

\bibitem{hb22} I. Sakalli, Phys. Rev. D 94, 084040 (2016).

\bibitem{hb23} P. Boonserm, T. Ngampitipan, and P. Wongjun, The European Physical Journal C 79 (2019).

\bibitem{hb24} A. Al-Badawi, I. Sakalli, and S. Kanzi, Annals of Physics 412, 168026 (2020).


\bibitem{hb25} A. Al-Badawi, S. Kanzi, and ˙I. Sakallı, The European Physical Journal Plus 135, 1 (2020).

\bibitem{hb26} S. Barman, The European Physical Journal C 80, 50 (2020).

\bibitem{hb27} M. Okyay and A. Ovgun, Journal of Cosmology and Astroparticle Physics, 2022, 009 (2022).

\bibitem{hb28} A. Al-Badawi, S. Kanzi, and ˙I. Sakallı, Annals of Physics 452, 169294 (2023).

\bibitem{hb29} M. Visser, Physical Review A 59, 427 (1999).

\bibitem{hb30} P. Boonserm and M. Visser, Physical Review D 78, 101502 (2008).

\bibitem{hb33} P. Boonserm, C.\thinspace H. Chen, T. Ngampitipan, and P. Wongjun
Phys. Rev. D 104, 084054 ]

\bibitem{qn28} S. W. Hawking, Nature 248, 30 (1974).

\bibitem{qn29} S. W. Hawking, Commun. Math. Phys. 43, 199 (1975), [Erratum:
Commun.Math.Phys. 46, 206 (1976).

\bibitem{gundlach1}Gundlach C, Price R H, and Pullin J 1994 Late time behavior of stellar collapse and explosions: 2. Nonlinear evolution \emph{Phys. Rev. D} \textbf{49} 890-899

\bibitem{yg2017}Y.-G. Miao and Z.-M. Xu, Hawking Radiation of Five-Dimensional Charged Black Holes with Scalar
Fields, Phys. Lett. B 772, 542 (2017).

\bibitem{fg2016} F. Gray, S. Schuster, A. Van–Brunt, and M. Visser, The Hawking Cascade from a Black Hole Is
Extremely Sparse, Class. Quantum Grav. 33, 115003 (2016).

\bibitem{ac2020} A. Chowdhury and N. Banerjee, Greybody Factor and Sparsity of Hawking Radiation from a Charged
Spherical Black Hole with Scalar Hair, Phys. Lett. B 805, 135417 (2020).

\bibitem{sh2016} S. Hod, The Hawking Cascades of Gravitons from Higher-Dimensional Schwarzschild Black Holes,
Phys. Lett. B 756, 133 (2016) [arXiv:1605.08440].

\bibitem{sh2015} S. Hod, The Hawking Evaporation Process of Rapidly-Rotating Black Holes: An Almost Continuous
Cascade of Gravitons, Eur. Phys. J. C 75, 329 (2015) [arXiv:1506.05457].

\end{thebibliography}
\end{document}